\documentclass[11pt]{article}
\usepackage[margin=1in]{geometry}
\usepackage{graphicx}
\usepackage{booktabs,longtable,array,calc}
\usepackage{amsmath,amssymb}
\usepackage{float}
\usepackage{tikz}
\usetikzlibrary{positioning,arrows.meta}
\usepackage{microtype}
\usepackage[hidelinks]{hyperref}
\usepackage{placeins}
\title{From Contact PPG to Camera-Derived rPPG: Property-Specific Recoverability, Dynamical Correspondence, and Fitzpatrick-Associated Heterogeneity}
\author{Timothy Oladunni$^{1,*}$ \and Farouk Ganiyu Adewumi$^{1}$\\[4pt]
\small $^{1}$Department of Computer Science, Morgan State University, Baltimore, Maryland, USA\\
\small $^{*}$Corresponding author: \href{mailto:timothy.oladunni@morgan.edu}{timothy.oladunni@morgan.edu}}
\date{}
\begin{document}
\maketitle
\section{Highlights}\label{highlights}

\textbullet{} A fixed camera-rPPG observation pathway reproduced the published MMPD CHROM correlation regime without post-hoc tuning, with modestly higher absolute error.

\textbullet{} Recoverability differed across endpoint, temporal, spectral, and
nonlinear signal properties.

\textbullet{} No detectable recording-specific PPG-rPPG maximal-Lyapunov correspondence was found (r\ensuremath{\lambda}=0.0231;
permutation p=0.5584).

\textbullet{} Observation context reduced subject-held-out PPG--rPPG HR MAE by up to 13.32\%, with motion and lighting providing the predictive gain.

\textbullet{} Fitzpatrick-associated endpoint heterogeneity persisted after SNR adjustment, while the tested dynamical discrepancy showed a different heterogeneity pattern.

\section{Abstract}\label{abstract}

Camera-derived remote photoplethysmography (rPPG) is commonly validated through endpoint accuracy, but endpoint performance alone does not establish whether other physiological properties of the source contact photoplethysmography (PPG) remain preserved recording by recording. We evaluated property-specific PPG-to-rPPG recoverability on 655 recordings from the Multi-Domain Mobile Video Physiology Dataset (MMPD) using the established CHROM method as a fixed, reproducible camera-rPPG observation pathway. The pathway reproduced the published MMPD CHROM correlation regime (heart-rate MAE 15.26 bpm; Pearson $r=0.0801$). Matched-versus-shuffled validation revealed modest recording-specific autocorrelation correspondence, whereas spectral and recurrence-rate measures showed little matched discrimination; determinism was saturated and therefore non-discriminative under the present parameterization. Maximal Lyapunov exponents showed essentially no recording-specific PPG--rPPG correspondence ($r_{\lambda}=0.0231$; condition-preserving permutation $p=0.5584$), despite population-level overlap. Endpoint discrepancy also exhibited Fitzpatrick-associated heterogeneity after adjustment for lighting and motion, including a Fitzpatrick VI versus III contrast of +9.32 bpm (95\% CI 3.78--14.86), while the tested dynamical discrepancy showed no corresponding gradient. Adding aggregate RGB signal-to-noise ratio (SNR) did not materially account for the endpoint contrast, and a simple Beer--Lambert attenuation model did not reproduce the measured SNR pattern. Finally, a subject-held-out condition-aware experiment showed that observation context provided incremental predictive information. Across linear, ridge, and random-forest learners, adding motion and lighting consistently reduced subject-held-out MAE relative to the corresponding rPPG-HR-only calibration baseline, with reductions of up to 13.32\%; adding Fitzpatrick group or aggregate SNR provided no further improvement. These findings support the conclusion that, under the evaluated observation pathway, recoverability is not a single global property of camera-derived rPPG: physiological properties can differ in recording-specific preservation and in their dependence on observation conditions, and population-level plausibility does not establish preservation of individual recordings.

Keywords: remote photoplethysmography; contact PPG; signal
recoverability; Lyapunov exponent; Fitzpatrick classification; signal-to-noise ratio

\section{Introduction}\label{introduction}

Remote photoplethysmography (rPPG) estimates cardiovascular pulsatility from subtle camera-recorded variations in skin reflectance and enables contactless physiological sensing for telemonitoring, home monitoring, human--computer interaction, and related applications \cite{mcduff2015survey}. Contact photoplethysmography (PPG), by contrast, measures cardiac-synchronous changes in peripheral blood volume through direct optical interaction with tissue \cite{allen2007ppg}. Camera-derived rPPG is therefore not a direct substitute for contact PPG: tissue optics, illumination, camera response, spatial sampling, compression, facial-region selection, and signal reconstruction transform the physiological observation before an endpoint is estimated \cite{wang2017algorithmic,mcduff2015survey,jacques2013optical}.

Most rPPG validation consequently emphasizes endpoint performance, particularly heart-rate (HR) mean absolute error (MAE), root-mean-square error (RMSE), percentage error, and correlation \cite{liu2023rppgtoolbox,xiao2024review,dasari2021bias}. These metrics are necessary but do not establish signal equivalence. A sensing transformation may preserve one physiological or signal property while degrading another. Moreover, similarity at the population level does not imply preservation of recording identity. For a property operator $\phi_j$, similarity between the marginal distributions of $\phi_j(X^{(p)})$ and $\phi_j(X^{(r)})$ does not establish that $\phi_j(x_i^{(p)})$ corresponds to $\phi_j(x_i^{(r)})$ for the same recording $i$. Thus, endpoint accuracy, signal equivalence, and property-specific recoverability are distinct questions.

This distinction motivates a property-specific view of PPG-to-rPPG validation. Under a camera observation pathway, some properties may remain recoverable while others may not. Conventional temporal, spectral, and recurrence structure provide complementary tests of waveform preservation, while nonlinear dynamics provide a more stringent test of whether recording-specific organization survives the observation transformation. State-space reconstruction provides a framework for characterizing dynamical structure from scalar observations \cite{takens1981detecting}; within the reconstructed trajectory, the maximal Lyapunov exponent ($\lambda_{\max}$) characterizes local exponential divergence and can be estimated from short time series using practical methods such as Rosenstein's algorithm \cite{rosenstein1993practical}. Prior rPPG work has examined dynamical reconstruction beyond endpoint HR \cite{he2022dynamics}, but estimating a dynamical quantity separately in PPG and rPPG does not by itself demonstrate preservation. Recording-specific correspondence must exceed what remains when the PPG--rPPG pairing is deliberately broken.

Recoverability may also depend on the observation conditions under which the camera signal is formed. Skin optical properties, illumination, motion, camera processing, and algorithmic assumptions can affect different components of rPPG \cite{jacques2013optical,dasari2021bias,wang2017algorithmic}. The Multi-Domain Mobile Video Physiology Dataset (MMPD) was designed to benchmark rPPG across variation in Fitzpatrick classification, lighting, and motion/activity \cite{tang2023mmpd}, making it suitable for asking whether observation-associated discrepancies are uniform across physiological properties. Fitzpatrick classification is treated here as a measured grouping variable rather than a direct measure of melanin or an identified causal optical mechanism \cite{weir2024skintone}. Likewise, aggregate signal-to-noise ratio (SNR) is evaluated as one possible quality descriptor rather than assumed to explain all downstream discrepancy \cite{elgendi2024sqi}.

The unresolved problem is therefore not merely whether camera-derived rPPG produces plausible HR estimates, but which properties of synchronized contact PPG retain recording-specific correspondence after camera observation and reconstruction, whether loss varies across measured observation conditions, and whether any of that context is actionable on unseen subjects. Existing rPPG studies provide endpoint benchmarks, signal-quality analyses, nonlinear characterization, and systematic evaluation frameworks \cite{mcduff2015survey,wang2017algorithmic,dasari2021bias,he2022dynamics,liu2023rppgtoolbox,xiao2024review}; however, these approaches do not by themselves establish matched preservation of multiple properties from the same synchronized contact-PPG recordings. We address this gap using a fixed CHROM-based camera-rPPG observation pathway, matched-versus-shuffled correspondence controls, subject-aware heterogeneous analyses, and subject-held-out prediction. The objective is to characterize property-specific and condition-dependent recoverability under the evaluated pathway, not to optimize a new rPPG estimator or claim general interchangeability between contact PPG and camera-derived rPPG.

\paragraph{Relation to Prior Representation Frameworks.}
The present study follows a progression from representation-level information to observation-level recoverability. Complementary Feature Domains (CFD) motivates the premise that physiological information can depend on the representation in which a signal is analyzed: distinct feature domains may retain different task-relevant information rather than serving as interchangeable descriptions of the same observation \cite{oladunni2025cfd}. Cardiac Stability Theory (CST) extends this representation-level perspective to cardiovascular signals by considering temporal and dynamical organization in contact PPG as physiologically informative structure beyond a single conventional endpoint \cite{oladunni2026cst}. Both perspectives therefore motivate asking not only what information is represented in an observed physiological signal, but what happens to that information when the observation mechanism itself changes.

The present work addresses that observation-level question. Let $x_i^{(p)}$ denote contact PPG from recording $i$, and represent the corresponding camera-derived signal as
\[
x_i^{(r)}=\mathcal{H}_{\theta_i}\!\left(x_i^{(p)}\right),
\]
where $\mathcal{H}_{\theta_i}$ denotes the camera observation and reconstruction pathway under recording-specific conditions $\theta_i$. For a physiological property operator $\phi_j$, the relevant question is whether the property extracted before observation, $\phi_j(x_i^{(p)})$, remains associated with the corresponding property after observation, $\phi_j(x_i^{(r)})$. Importantly, similarity of the population-level distributions,
\[
P\!\left[\phi_j(X^{(p)})\right] \approx P\!\left[\phi_j(X^{(r)})\right],
\]
does not establish recording-specific preservation $\phi_j(x_i^{(p)})\leftrightarrow\phi_j(x_i^{(r)})$. This motivates the property-specific recoverability framework evaluated here: endpoint, conventional structural, and nonlinear dynamical properties are assessed separately across the contact-PPG--to--rPPG observation boundary.

The relationship to CFD and CST is therefore conceptual rather than assumptive. The present analysis does not require either framework to hold and does not constitute a direct test of either theory. Instead, it examines the observation-level question that follows naturally from their representation-centered perspective: when a physiologically informative signal representation is subjected to a different sensing and reconstruction process, which of its properties remain recoverable for the same recording?

\begin{figure}[!htbp]
\centering
\resizebox{0.98\textwidth}{!}{\begin{tikzpicture}[
  >=Latex,
  font=\small,
  box/.style={draw, rounded corners=1.5pt, align=center, inner sep=8pt, minimum height=1.18cm},
  main/.style={box, fill=black!3},
  prop/.style={box, fill=black!1, minimum width=4.05cm},
  arrow/.style={->, line width=0.65pt}
]

\node[main, text width=3.2cm] (ppg) at (0,0)
  {\textbf{Contact PPG}\\[2pt] $x_i^{(p)}(t)$};

\node[main, text width=5.0cm] (obs) at (6.0,0)
  {\textbf{Camera observation pathway}\\[2pt]
   $x_i^{(r)}=\mathcal H_{\theta_i}\!\left(x_i^{(p)}\right)$\\[-1pt]
   {\footnotesize observation conditions $\theta_i$}};

\node[main, text width=3.2cm] (rppg) at (12.0,0)
  {\textbf{Camera-derived rPPG}\\[2pt] $x_i^{(r)}(t)$};

\draw[arrow] (ppg) -- (obs);
\draw[arrow] (obs) -- (rppg);

\node[font=\bfseries] (propertylabel) at (6.0,-1.60)
  {Evaluate recoverability property by property};

\node[prop] (hr) at (1.55,-2.95)
  {\textbf{Endpoint}\\ $\phi_{HR}(x)$};
\node[prop] (struct) at (6.0,-2.95)
  {\textbf{Conventional structure}\\ $\phi_{S}(x)$};
\node[prop] (dyn) at (10.45,-2.95)
  {\textbf{Nonlinear dynamics}\\ $\phi_{\lambda}(x)=\lambda_{\max}(x)$};

\draw[line width=0.5pt] (1.55,-2.10) -- (10.45,-2.10);
\draw[arrow] (6.0,-1.83) -- (6.0,-2.10);
\draw[arrow] (1.55,-2.10) -- (hr.north);
\draw[arrow] (6.0,-2.10) -- (struct.north);
\draw[arrow] (10.45,-2.10) -- (dyn.north);

\node[main, text width=5.35cm] (marg) at (2.75,-5.05)
  {\textbf{Marginal plausibility}\\[3pt]
   $P\!\left[\phi_j(X^{(p)})\right]\approx
    P\!\left[\phi_j(X^{(r)})\right]$};

\node[main, text width=5.35cm] (pair) at (9.25,-5.05)
  {\textbf{Recording-specific correspondence}\\[3pt]
   $\phi_j(x_i^{(p)})\leftrightarrow\phi_j(x_i^{(r)})$};

\node[font=\Large] at (6.0,-5.05) {$\not\Rightarrow$};

\end{tikzpicture}}
\caption{\textbf{Conceptual Framework for Property-Specific PPG-to-rPPG Recoverability.} \textit{Contact PPG is transformed through a heterogeneous camera observation pathway, and recoverability is evaluated separately for physiological properties. Population-level plausibility does not by itself establish recording-specific correspondence.}}
\label{fig:conceptual_recoverability}
\end{figure}
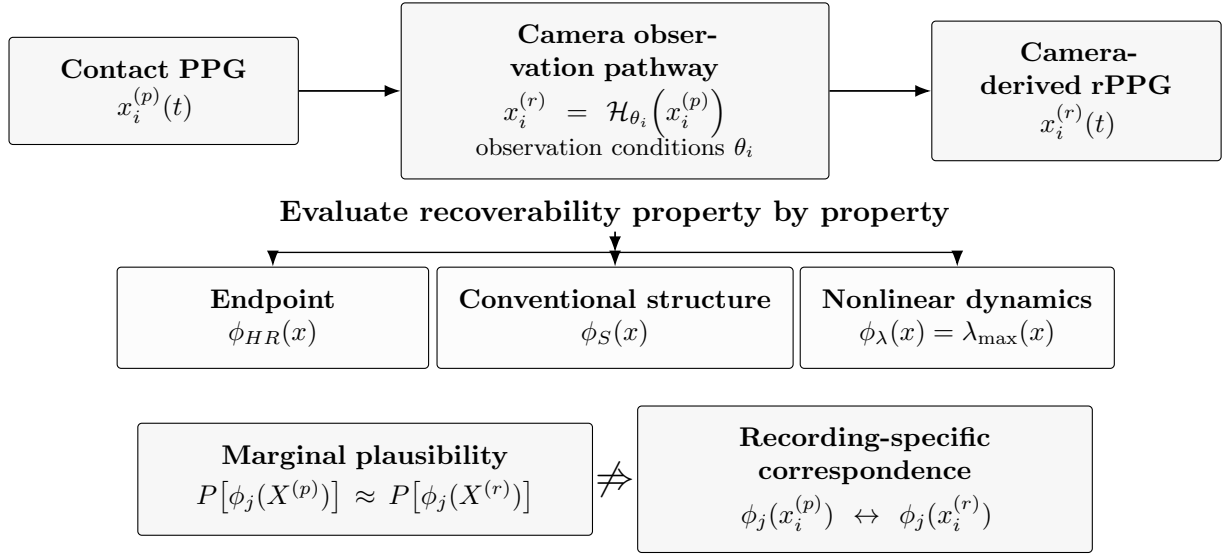

Prior work investigated topological signal representations as a framework for characterizing and mitigating skin-tone-associated effects in optical physiological measurements \cite{oladunni2026skin}. The present study addresses a distinct question: whether specific physiological and dynamical properties of a contact-PPG source remain recoverable after transformation through a camera-rPPG observation pathway. Figure~\ref{fig:conceptual_recoverability} summarizes the study's property-specific view of PPG-to-rPPG recoverability and distinguishes the source properties, camera observation pathway, and corresponding recovered properties.

\subsection{Research questions and empirical mapping}\label{research-questions-and-empirical-mapping}

To make the study logic explicit, Table~\ref{tab:rqmap} maps each research question to its empirical test and the result that answers it. The table is an organizational synthesis of the analyses reported below and introduces no new analysis.

\begingroup
\small
\setlength{\tabcolsep}{3pt}
\renewcommand{\arraystretch}{1.12}
\begin{longtable}{@{}p{0.06\textwidth}p{0.25\textwidth}p{0.25\textwidth}p{0.36\textwidth}@{}}

\multicolumn{4}{@{}p{\linewidth}@{}}{\textbf{Table~\thetable{} \textbar{} Research Questions, Empirical Tests, and Principal Findings}\par\vspace{2pt}\textit{Each row maps a research question to the mathematical or statistical analysis used to evaluate it and the principal empirical conclusion. Detailed effect estimates and uncertainty measures are reported in the corresponding Results tables.}} \\[4pt]
\toprule
\textbf{RQ} & \textbf{Research question} & \textbf{Empirical test} & \textbf{Empirical answer} \\
\midrule
\endfirsthead
\toprule
\textbf{RQ} & \textbf{Research question} & \textbf{Empirical test} & \textbf{Empirical answer} \\
\midrule
\endhead
\bottomrule
\endlastfoot
\label{tab:rqmap}RQ1 & How accurately is heart rate recovered from contact PPG after transformation to camera-derived rPPG under the fixed observation pathway? &
Endpoint discrepancy $D_{HR}=|HR^{(p)}-HR^{(r)}|$ summarized by MAE, RMSE, MAPE, signed error, and Pearson correlation. &
\textbf{Limited HR recovery under the evaluated observation pathway.} MAE=15.26 bpm and $r=0.0801$, indicating substantial endpoint error and weak recording-level association (Table~\ref{tab:chrom}). \\

RQ2 & Which conventional temporal, spectral, and recurrence properties retain recording-specific correspondence between matched PPG and rPPG? &
Property-specific matched-versus-shuffled structural correspondence for ACF similarity, cardiac-band PSD distance, and recurrence-rate difference (Algorithm~1); determinism assessed for discriminative variability. &
\textbf{Selective recording-specific structural correspondence.} ACF showed a modest matched advantage, whereas PSD and recurrence-rate differences provided little matched-versus-shuffled separation. Determinism was non-discriminative under the implemented parameterization (Table~\ref{tab:structural}). \\

RQ3 & Is maximal Lyapunov exponent ($\lambda_{\max}$) preserved as a recording-specific dynamical property across matched PPG--rPPG recordings? &
Study-specific $\hat{\lambda}_{\max}$ estimation and paired discrepancy (Algorithm~2), followed by recording-specific correspondence and a 10,000-permutation condition-preserving subject-permutation null test (Algorithm~1). &
\textbf{No detectable recording-specific dynamical correspondence.} Despite population-level dynamical plausibility, matched PPG--rPPG estimates were essentially uncorrelated ($r_{\lambda}=0.0231$), and the observed association was not distinguishable from the condition-preserving subject-permutation null (Table~\ref{tab:lyapunov}). \\

RQ4 & Which measured observation factors are associated with endpoint and dynamical discrepancies, and are the heterogeneity patterns property-specific? &
Subject-aware mixed-effects models for $D_{HR}$ and $D_{\lambda}$, with Fitzpatrick group, lighting, motion/activity, interaction block tests, and aggregate RGB-SNR sensitivity analysis (Algorithm~3). &
\textbf{Observation-associated discrepancy was property-specific.} $D_{HR}$ varied substantially with motion/activity and Fitzpatrick group, including evidence of a Fitzpatrick$\times$motion interaction, whereas $D_{\lambda}$ showed a different heterogeneity pattern with little Fitzpatrick- or lighting-associated variation. Aggregate RGB SNR did not materially account for the endpoint heterogeneity (Tables~\ref{tab:model1}--\ref{tab:model1b}). \\

RQ5 & To what extent can measured observation context reduce the PPG--rPPG endpoint discrepancy, as measured by MAE, on unseen subjects? &
Subject-disjoint nested-information prediction with linear, ridge, and random-forest learners; out-of-fold MAE and subject-level bootstrap $\Delta MAE$ relative to the rPPG-HR-only calibration model (Algorithm~4). &
\textbf{Partially.} Motion and lighting reduced subject-held-out MAE beyond rPPG-HR-only calibration across linear, ridge, and random-forest learners. At $M_2$, MAE was reduced by 13.32\%, 13.32\%, and 12.77\%, respectively, relative to each learner's $M_0$ baseline. Adding Fitzpatrick group or aggregate RGB SNR provided no further improvement, and residual error remained substantial (Figure~\ref{fig:gapreduction}). \\

\end{longtable}
\endgroup

\subsection{Contributions}\label{contributions}

This study makes four principal contributions.
\begin{enumerate}
\item \textbf{Property-specific formulation of physiological recoverability.}
We formulate PPG-to-rPPG validation as a property-specific problem rather than a single modality-level fidelity judgment, separating endpoint, conventional structural, and nonlinear dynamical properties.

\item \textbf{Recording-specific correspondence as a criterion distinct from marginal plausibility.}
For structural and dynamical properties, matched PPG--rPPG comparisons are evaluated against correspondence-breaking shuffled controls so that population-level similarity is not interpreted as preservation of individual recordings.

\item \textbf{Subject-aware characterization of property-specific heterogeneity.}
We quantify endpoint and dynamical discrepancies across Fitzpatrick group, lighting, and motion/activity using repeated-measures models, interaction tests, and an aggregate RGB-SNR sensitivity analysis to determine whether heterogeneous observation conditions affect different properties in the same way.

\item \textbf{Subject-held-out test of whether observation context is actionable.}
Using nested information sets and simple linear, ridge, and random-forest learners, we test whether measured observation context reduces the PPG--rPPG HR gap on unseen subjects beyond an rPPG-HR-only calibration baseline.
\end{enumerate}

Together, these contributions provide a unified test of what survives the PPG-to-rPPG observation pathway, whether preservation is recording-specific, where discrepancies vary, and whether measured observation context can partially narrow the endpoint gap.

\section{Methods}\label{methods}

\subsection{Study design and analysis
sequence}\label{study-design-and-prespecified-analysis-sequence}

This was a retrospective, recording-matched analysis of synchronized
contact PPG and facial video from the Multi-Domain Mobile Video
Physiology Dataset (MMPD). The analysis followed a fixed hierarchy rather
than an algorithm-selection exercise. First, we established a fixed
camera-rPPG observation pathway and verified its benchmark behavior.
Second, we quantified endpoint heart-rate discrepancy. Third, we tested
conventional temporal and spectral structure against shuffled-pair
negative controls. Fourth, we applied the same study-specific
maximal-Lyapunov estimator to PPG and rPPG and tested recording-specific
correspondence. Fifth, we quantified heterogeneity of endpoint and
dynamical discrepancies across Fitzpatrick group, lighting, and motion
using subject-aware mixed-effects models. Finally, we evaluated measured
RGB SNR against a simple optical-attenuation model, examined SNR-adjusted
endpoint heterogeneity, and tested whether measured observation context
reduced held-out endpoint error. No Fitzpatrick-specific tuning,
alternative rPPG-method search, or post-hoc selection of SNR definitions
was performed.

The property-specific quantities, recording-correspondence criterion, and heterogeneous-observation model used throughout the analysis are formalized in Section~\ref{mathematical-formulation}.

The unit of analysis was the recording, while repeated recordings from
the same participant were handled through subject-level random
intercepts in inferential models. The expected MMPD design comprised 33
subjects with 20 recordings each (660 expected recordings). Five source
recordings were absent from the available local dataset, yielding 655
available recordings for the fixed-pathway benchmark. All exclusions and
estimator failures in downstream dynamical analysis followed the stated
validity rules and were not determined by observed cross-modal agreement.

\subsection{Mathematical formulation of property-specific recoverability}
\label{mathematical-formulation}

The analysis is organized around a latent physiological process observed through two sensing pathways. Let $s_i(t)$ denote the underlying cardiovascular process for recording $i$. The synchronized contact PPG and camera pathway are represented as
\begin{equation}
x_i^{(p)} = \mathcal{P}\!\left[s_i\right],
\qquad
v_i = \mathcal{V}_{\theta_i}\!\left[s_i\right],
\qquad
x_i^{(r)} = \mathcal{O}_{\phi}(v_i),
\label{eq:observation}
\end{equation}
where $x_i^{(p)}$ is the reference contact-PPG waveform, $v_i$ is the facial video, $\mathcal{V}_{\theta_i}$ denotes the camera observation process under measured condition vector $\theta_i$, and $\mathcal{O}_{\phi}$ is a fixed camera-rPPG observation pathway that produces $x_i^{(r)}$. In this experiment, $\mathcal{O}_{\phi}=\mathcal{O}_{\mathrm{CHROM}}$ using the established CHROM method \cite{dehaan2013robust}. In the present dataset, $\theta_i$ includes Fitzpatrick group, lighting, and motion/activity descriptors. Equation~\ref{eq:observation} is an organizational observation model, not an identified causal optical model and not an assumption that an exact inverse from rPPG to PPG exists.

Let $\phi_j$ denote a physiological or signal property indexed by $j$. The property values extracted from the two observations are
\begin{equation}
z_{ij}^{(p)}=\phi_j\!\left(x_i^{(p)}\right),
\qquad
z_{ij}^{(r)}=\phi_j\!\left(x_i^{(r)}\right),
\qquad
j\in\mathcal{J},
\label{eq:propertymap}
\end{equation}
with the evaluated property family
\begin{equation}
\mathcal{J}
=
\left\{
HR,\,
ACF,\,
PSD,\,
RQA,\,
\lambda_{\max}
\right\}.
\label{eq:propertyfamily}
\end{equation}
The notation is deliberately property-specific: the study does not assume that successful recovery of one member of $\mathcal{J}$ implies recovery of another.

For properties naturally compared by a distance or absolute difference, recording-level recoverability loss is written generically as
\begin{equation}
D_{ij}
=
d_j\!\left(
z_{ij}^{(p)},z_{ij}^{(r)}
\right).
\label{eq:propertyloss}
\end{equation}
The two principal scalar discrepancies are therefore
\begin{equation}
D_{i,HR}
=
\left|
HR_{\mathrm{rPPG},i}
-
HR_{\mathrm{PPG},i}
\right|,
\qquad
D_{i,\lambda}
=
\left|
\lambda_{\max,i}^{(r)}
-
\lambda_{\max,i}^{(p)}
\right|.
\label{eq:principaldiscrepancies}
\end{equation}
For ACF, PSD, and recurrence analyses, $d_j$ is replaced by the corresponding similarity or distance statistic defined below.

A second requirement is \emph{recording specificity}. Similar marginal distributions alone do not establish that the property belonging to recording $i$ survives the observation pathway. Let $\pi(i)$ denote a correspondence-breaking permutation that maps a source PPG recording to a nonmatched rPPG recording while preserving the specified condition structure. For a similarity statistic $q_j$, evidence of recording-specific preservation is supported when
\begin{equation}
\mathbb{E}\!\left[
q_j\!\left(z_{ij}^{(p)},z_{ij}^{(r)}\right)
\right]
>
\mathbb{E}\!\left[
q_j\!\left(z_{ij}^{(p)},z_{\pi(i)j}^{(r)}\right)
\right],
\label{eq:matchedcriterion}
\end{equation}
with the direction reversed for a distance statistic. For $\lambda_{\max}$, the analogous test is whether the matched recording correlation exceeds its condition-preserving subject-permutation null. Equation~\ref{eq:matchedcriterion} formalizes why the manuscript distinguishes population-level similarity from recording-specific recoverability.

Finally, heterogeneity is modeled as variation in the property-specific discrepancy conditional on measured observation factors. For property $j$, the subject-aware model is
\begin{equation}
D_{ij}
=
\beta_{0j}
+
\boldsymbol{\beta}_{F,j}^{\top}F_i
+
\boldsymbol{\beta}_{L,j}^{\top}L_i
+
\boldsymbol{\beta}_{M,j}^{\top}M_i
+
u_{s(i),j}
+
\varepsilon_{ij},
\label{eq:mixedmodelgeneral}
\end{equation}
where $F_i$, $L_i$, and $M_i$ encode Fitzpatrick group, lighting, and motion/activity; $u_{s(i),j}$ is a subject-specific random intercept; and $\varepsilon_{ij}$ is the recording-level residual. The fitted models in this study instantiate Equation~\ref{eq:mixedmodelgeneral} separately for $D_{HR}$ and $D_{\lambda}$. Thus, Fitzpatrick group, lighting, and motion/activity enter as parallel explanatory factors rather than as definitions of recoverability.

This formulation separates three questions that are otherwise easily conflated: whether a property can be estimated from rPPG, whether its recording identity is preserved relative to the synchronized PPG, and whether its loss changes across observation conditions. Algorithms~\ref{alg:recoverability}--\ref{alg:heterogeneous} operationalize the property-specific characterization framework, while Algorithm~\ref{alg:conditionaware} separately tests whether measured observation context can translate the identified heterogeneity into improved held-out endpoint prediction.

\subsection{Dataset, recording structure, and
metadata}\label{dataset-recording-structure-and-metadata}

Each available MMPD record contained facial RGB video and synchronized
ground-truth PPG. Video consisted of 1,800 frames at 30 frames/s,
corresponding to approximately 60 s per recording. The analysis retained
the dataset\textquotesingle s heterogeneous acquisition labels,
including Fitzpatrick group (III-VI), lighting condition, and
motion/activity condition. These variables were treated as observed
acquisition descriptors rather than manipulated causal exposures.

Raw inspection identified inconsistent Fitzpatrick metadata for Subject
24: one recording was labeled Fitzpatrick IV and the remaining
recordings were labeled Fitzpatrick III. To avoid assigning a post-hoc
skin-type label, Subject 24 was excluded only from Fitzpatrick-dependent
analyses and retained for analyses not requiring skin-type
classification. After removal of one duplicated matched key, the
heterogeneous-analysis table contained 645 unique matched records; the
Fitzpatrick analysis population contained 625 recordings from 32
subjects.

\subsection{Fixed camera-to-rPPG observation operator}\label{standalone-chrom-observation-pipeline}

Camera-derived rPPG was generated with a fixed observation operator, $\mathcal{O}_{\phi}$, instantiated in this experiment by the established chrominance-based CHROM method \cite{dehaan2013robust}. CHROM is not a methodological contribution of this study and was not optimized against the downstream recoverability results. It was selected as a reproducible, unsupervised camera-to-rPPG operator and held fixed across structural, dynamical, heterogeneous, and condition-aware analyses.

For reproducibility, facial ROIs were fixed within each recording, resized to $72\times72$ pixels, and reduced to frame-wise RGB spatial means. The standard CHROM construction was applied in 1.6-s windows with 50\% overlap, channel normalization, chrominance projection, bandpass filtering, adaptive scaling, and overlap-add reconstruction \cite{dehaan2013robust}. Benchmark HR estimation used the same detrending, 0.6--3.3-Hz bandpass, and frequency-domain peak-estimation pathway for all recordings. Benchmark results are reported in Section~\ref{dataset-accounting-and-frozen-chrom-benchmark}. Thereafter, the scientific analyses treat $x^{(r)}=\mathcal{O}_{\phi}(v)$ as the camera-derived observation; conclusions concern recoverability under this fixed operator rather than CHROM development itself.

\subsection{Endpoint heart-rate
recoverability}\label{endpoint-heart-rate-recoverability}

Endpoint and structural recoverability were evaluated according to Algorithm~\ref{alg:recoverability}.

Reference heart rate, $HR_{\mathrm{PPG}}$, was derived from the synchronized contact PPG, whereas camera-derived heart rate, $HR_{\mathrm{rPPG}}$, was estimated from the fixed rPPG waveform using the benchmark evaluation procedure. For each recording, signed endpoint error was $HR_{\mathrm{rPPG}}-HR_{\mathrm{PPG}}$ and absolute endpoint discrepancy was $D_{HR}=|HR_{\mathrm{rPPG}}-HR_{\mathrm{PPG}}|$. In the reproducibility code, these quantities are stored as \texttt{HR\_CHROM} and \texttt{HR\_GT}, respectively. The primary
heterogeneous-observation outcome was DHR because it measures the
magnitude of endpoint discrepancy without cancellation of positive
and negative differences. No recording was removed because its HR discrepancy was
large.

Overall endpoint performance was summarized using MAE, RMSE, MAPE,
signed error, and Pearson correlation. These metrics were used as
endpoint-level evidence only; they were not interpreted as measures of
waveform or nonlinear-structure preservation.

\subsection{Conventional temporal and spectral
structure}\label{conventional-temporal-and-spectral-structure}

To test whether weak endpoint agreement implied complete loss of
lower-level structure, we compared matched PPG-rPPG recordings using
temporal and spectral measures. Autocorrelation-function (ACF)
similarity quantified similarity of temporal repetition structure.
Cardiac-band power-spectral-density (PSD) distance quantified
differences in normalized spectral organization over the prespecified
physiological band. Recurrence-rate discrepancy was evaluated as an
additional state-space summary.

Each structural metric was subjected to a matched-versus-shuffled
negative control. The matched analysis used the synchronized PPG and
rPPG from the same recording. The shuffled analysis deliberately broke
recording correspondence while preserving the collection of signals. A
structural measure was considered evidence of recording specificity only
when matched pairs were more similar than shuffled pairs in the expected
direction. This control was essential because physiological recordings
can share generic cardiac-band structure even when they do not preserve
recording identity.

An initially evaluated determinism statistic was identically 1.0 across
the analyzed signals. Because a constant statistic cannot discriminate
matched from mismatched recordings, it was classified as a metric
failure and excluded from evidentiary claims rather than interpreted as
perfect dynamical preservation.

\subsection{Study-specific maximal-Lyapunov
estimator}\label{frozen-maximal-lyapunov-estimator}

The estimator, validity rules, and recording-level matching procedure are specified in Algorithm~\ref{alg:lyapunov}.

Maximal Lyapunov exponent (\ensuremath{\lambda}max) was used as the nonlinear
property for the PPG-to-rPPG correspondence test. The same estimator and
validity criteria were applied to reference PPG and fixed camera-derived
rPPG; estimator settings were not adapted according to the observed
cross-modal agreement.

Signals entered a common analysis pathway and were resampled to 100 Hz
so that delay, Theiler exclusion, and divergence time were expressed on
the same time base for PPG and rPPG. A 20-s analysis segment was used.
The delay tau was estimated from average mutual information with a
maximum search of 50 samples. State-space reconstruction used embedding
dimension m=5. Temporally adjacent points were excluded from
nearest-neighbor selection using a 100-ms Theiler window. Lambda\_max
was then estimated from the approximately linear region of the mean
log-divergence curve using a Rosenstein-style procedure.

A recording could fail because of
the stated signal-quality gate, failure to estimate tau,
insufficient embedding trajectory, failure of the divergence fit, or a
non-finite/non-positive exponent. Failure status was recorded rather
than replaced or imputed. The primary cross-observation population was
the exact intersection of recordings with valid PPG and valid rPPG \ensuremath{\lambda}max
estimates. Consequently, the correlation analysis did not use
modality-specific imputation or selective deletion based on agreement.

\subsection{Recording-specific Lyapunov correspondence and permutation
test}\label{recording-specific-lyapunov-correspondence-and-permutation-test}

Recording-specific dynamical recoverability was quantified by Pearson
correlation between \ensuremath{\lambda}max estimated from the reference PPG and \ensuremath{\lambda}max
estimated from the matched rPPG. The primary effect was r\ensuremath{\lambda}, accompanied
by a 95\% confidence interval and conventional two-sided significance
test.

Because both signals can share condition-dependent population structure
without preserving recording identity, a subject-level
condition-preserving permutation test was used as the principal negative
control. Across 10,000 permutations, subject correspondence between PPG
and rPPG was broken while acquisition/recording condition structure was
preserved. The empirical two-sided probability was calculated with the
finite-sample correction p\_perm = (1 + number of
\textbar r\_perm\textbar{} \textgreater= \textbar r\_obs\textbar)/(B +
1), where B=10,000. The prespecified interpretation required the matched
r\ensuremath{\lambda} to exceed what could be obtained after breaking subject
correspondence; similarity of marginal \ensuremath{\lambda}max distributions alone was not
considered evidence of preservation.

\subsection{Construction of heterogeneous-recoverability
outcomes}\label{construction-of-heterogeneous-recoverability-outcomes}

For each deduplicated matched recording, two absolute discrepancies were
defined: $D_{HR}=|HR_{\mathrm{rPPG}}-HR_{\mathrm{PPG}}|$ and D\ensuremath{\lambda} =
\textbar \ensuremath{\lambda}max,PPG - \ensuremath{\lambda}max,rPPG\textbar. DHR measures endpoint loss,
whereas D\ensuremath{\lambda} measures numerical disagreement in the selected nonlinear
property. D\ensuremath{\lambda} was not treated as evidence of preserved dynamics unless
recording-specific correspondence was independently demonstrated by the
r\ensuremath{\lambda} analysis.

Descriptive summaries were computed by Fitzpatrick group before
adjustment. Subject 24 was excluded from all models containing
Fitzpatrick because of inconsistent metadata, leaving 625 recordings
from 32 subjects.

\subsection{Mixed-effects models for Fitzpatrick, lighting, and
motion}\label{mixed-effects-models-for-fitzpatrick-lighting-and-motion}

Two primary mixed-effects models were fitted as specified in Algorithm~\ref{alg:heterogeneous}. Model 1 used DHR as the
dependent variable; Model 2 used D\ensuremath{\lambda}. Both models included Fitzpatrick group, lighting condition, and motion/activity condition as fixed
effects and a random intercept for subject to account for repeated
recordings. Fitzpatrick III, incandescent lighting, and rotation were
reference categories. Models were fitted by maximum likelihood using
statsmodels MixedLM.

Fixed-effect coefficients represent adjusted mean differences in the
outcome relative to the reference category while holding the other
included acquisition variables constant. Ninety-five percent confidence
intervals and p-values were taken from the fitted mixed-effects model
output. Because the dataset contains repeated observations but
relatively few independent subjects in some Fitzpatrick strata,
inferential interpretation emphasizes effect estimates and confidence
intervals rather than treating the number of recordings as the number of
independent participants.

The primary models were additive. To test compounded heterogeneity directly rather than infer it from stratified cell means, two secondary interaction models were fitted for each outcome: Fitzpatrick$\times$motion and Fitzpatrick$\times$lighting, retaining the same subject random intercept and the four Fitzpatrick categories. Joint Wald tests evaluated the full interaction coefficient blocks (12 degrees of freedom for motion and 9 for lighting). Interaction interpretation was based on the block test rather than isolated cell coefficients. Subject 24 remained excluded because of inconsistent Fitzpatrick metadata. These observational interactions are not interpreted as causal optical mechanisms.

Fitzpatrick III-IV-V-VI categories were not collapsed into a binary light/dark variable.

\subsection{Subject-held-out condition-aware endpoint-discrepancy experiment}\label{condition-aware-gap-methods}

A follow-up experiment tested the extent to which observation variables identified in the heterogeneous analysis could reduce the endpoint discrepancy on unseen subjects, with out-of-fold MAE as the primary outcome. The complete computational sequence is summarized in Algorithm~\ref{alg:conditionaware}. The target was contact-PPG reference HR, and the base predictor was camera-derived rPPG HR. Here, $M_0$--$M_4$ denote nested predictor information sets rather than different learning algorithms; each information set was evaluated independently using linear regression, ridge regression, and random forest regression. Five nested specifications were evaluated: $M_0$ used rPPG HR alone; $M_1$ added motion/activity; $M_2$ added lighting; $M_3$ additionally added Fitzpatrick group; and $M_4$ additionally added aggregate RGB SNR. Subject 24 remained excluded, leaving 625 recordings from 32 subjects.

Generalization was evaluated with five-fold GroupKFold cross-validation at the subject level so that all recordings from a participant appeared entirely in either training or test data within a fold. Three deliberately simple learner families were compared: ordinary linear regression, ridge regression ($\alpha=1$), and random forest regression. Categorical variables were one-hot encoded within the training pipeline; numeric variables were standardized for the linear and ridge models. No deep neural network or end-to-end video model was used because the purpose was to test whether measured observation context reduced subject-held-out endpoint discrepancy, as measured primarily by MAE, rather than to maximize benchmark performance.

The primary outcome was out-of-fold MAE against the synchronized contact-PPG HR. RMSE, median absolute error, Pearson correlation, and $R^2$ were secondary diagnostic summaries and were not used to define the primary RQ5 comparison. Incremental improvement was evaluated relative to $M_0$ within each learner family. To respect subject dependence, the MAE difference between $M_0$ and each expanded model was assessed with a 5,000-resample subject-level bootstrap; a confidence interval entirely below zero was interpreted as evidence that the expanded model reduced MAE. This experiment was analyzed as a follow-up test of actionability, not as part of the original fixed-pathway benchmark.

\subsection{Measured RGB SNR}\label{measured-rgb-snr}

Recording-level red-, green-, and blue-channel SNR values were computed
from the extracted color traces and retained for all 655 available
recordings. The three channel SNRs were first summarized separately by
Fitzpatrick group to determine whether aggregate measured signal quality
showed the large monotonic decline that would be expected under a simple
attenuation-only explanation.

For the final sensitivity model, one aggregate SNR variable was
defined before fitting: SNRRGB = (SNR\_R + SNR\_G + SNR\_B)/3. No
alternative definitions such as minimum channel SNR, maximum channel
SNR, individual-channel model selection, or post-hoc weighted
combinations were searched.

\subsection{Simple optical-attenuation adequacy
check}\label{simple-optical-attenuation-adequacy-check}

A Beer-Lambert-inspired attenuation model was evaluated only as a
deliberately simplified mechanistic adequacy check. The model used
SNR\_dB = baseline\_SNR - 20 \ensuremath{\times} OD\_melanin to generate a monotonic
attenuation pattern across Fitzpatrick groups. Its purpose was not to
estimate melanin concentration or to provide a complete tissue-optics
model; rather, it tested whether a simple aggregate attenuation
explanation could reproduce the measured RGB SNR pattern.

Observed and predicted group-level SNR values were compared and
summarized by RMSE. Because the predicted decline was much larger than
the measured Fitzpatrick variation, the model was rejected as an
adequate explanation. No coefficients were subsequently tuned to force
agreement with the empirical data, and the model was not used to correct
HR or rPPG waveforms.

\subsection{SNR-adjusted sensitivity
model}\label{final-prespecified-snr-adjusted-sensitivity-model}

The mixed-effects heterogeneity models and SNR-adjusted sensitivity analysis are summarized in Algorithm~\ref{alg:heterogeneous}.

Model 1b tested whether measured aggregate RGB SNR statistically
accounted for the Fitzpatrick-associated endpoint coefficient. It
retained the exact Model 1 structure and added only SNRRGB: DHR
\textasciitilde{} Fitzpatrick + Lighting + Motion + SNRRGB +
(1\textbar subject). The analysis population, reference categories,
maximum-likelihood fitting procedure, and subject random intercept were
unchanged.

The principal quantities were the SNRRGB coefficient and the change in
the Fitzpatrick VI-versus-III coefficient relative to Model 1.
Coefficient attenuation was calculated descriptively as 100 \ensuremath{\times}
(\ensuremath{\beta}VI,Model1 - \ensuremath{\beta}VI,Model1b)/\ensuremath{\beta}VI,Model1. This quantity was explicitly not
interpreted as causal mediation because the analysis was observational
and did not satisfy a formal mediation design.

\subsection{Software, reproducibility, and analysis
freeze}\label{software-reproducibility-and-analysis-freeze}

Video processing, signal analysis, nonlinear estimation, statistical
modeling, and figure generation were implemented in Python. For the
maximal-Lyapunov analysis, the authoritative implementation was the
corrected prospective/frozen pipeline: all 33 subjects were retained for
dynamical analysis, recording indices were 0--19, and every expected
subject-recording combination was prospectively assigned a success or
failure status. The rPPG analysis applied the same frozen estimator and
parameters used for contact PPG; only waveform input/output handling
differed. The reproducibility record includes the fixed rPPG observation implementation, frozen rPPG waveforms, prospective PPG and rPPG
maximal-Lyapunov outputs, failure/accounting tables, matched structural
analyses, mixed-effects models, RGB SNR analysis, optical-attenuation
adequacy check, Model 1b sensitivity analysis, Fitzpatrick-by-condition interaction tests, the subject-held-out condition-aware gap-reduction experiment, and scripts that regenerate
the principal figures from frozen result files.

The primary evidence sequence was frozen through Model 1b. Two follow-up analyses were then added as explicitly labeled extensions rather than used to redefine the frozen primary results: (i) joint Fitzpatrick-by-motion and Fitzpatrick-by-lighting interaction tests, and (ii) a subject-held-out condition-aware endpoint learner using the already measured motion, lighting, Fitzpatrick, and aggregate-SNR variables. We did not proceed to Fitzpatrick-specific correction, alternative rPPG algorithms, channel-specific model searching, additional nonlinear metrics selected after observing \ensuremath{\lambda}max results, or open-ended covariate searching to reduce the Fitzpatrick coefficient.

\section{Algorithmic specification}\label{algorithmic-specification}

The following algorithms express the reported analyses at the level of their mathematical operations. They define the quantities a reader must reproduce while leaving software-specific details to the corresponding Methods subsections. The fixed CHROM observation operator is documented in Section~\ref{standalone-chrom-observation-pipeline}; no algorithm below changes that observation pathway or introduces post-hoc optimization.

\newcounter{manuscriptalgorithm}
\refstepcounter{manuscriptalgorithm}\label{alg:recoverability}
\textbf{Algorithm~\themanuscriptalgorithm{} \textbar{} Property-specific PPG--rPPG recoverability assessment}

\noindent\textit{Mathematical procedure for evaluating endpoint discrepancy and recording-specific structural and dynamical correspondence between synchronized contact PPG and camera-derived rPPG.}

\par\smallskip\noindent\rule{\linewidth}{0.4pt}\par\smallskip
\textbf{Input:} synchronized pairs $\mathcal{X}=\{(x_i^{P},x_i^{R})\}_{i=1}^{N}$, where $P$ denotes contact PPG and $R$ denotes camera-derived rPPG; property operators $\Phi=\{\phi_j\}_{j=1}^{J}$.

1. For each property $\phi_j$, map every synchronized recording pair into the corresponding property space,
\[
z_{ij}^{P}=\phi_j(x_i^{P}),\qquad z_{ij}^{R}=\phi_j(x_i^{R}).
\]
The operator $\phi_j$ may return a scalar endpoint, a structural descriptor, a recurrence statistic, or a nonlinear dynamical quantity.

2. For scalar endpoint HR, define the recording-level loss
\[
D_{HR,i}=\left|HR_i^{R}-HR_i^{P}\right|.
\]
For each conventional structural property $j\in\{ACF,PSD,REC\}$, compute the implemented matched-pair comparison functional
\[
S_{ij}^{\mathrm{match}}=\mathcal{S}_j(z_{ij}^{P},z_{ij}^{R}),
\]
where $\mathcal{S}_j$ denotes ACF similarity, cardiac-band PSD distance, or recurrence-rate discrepancy as defined in the corresponding Methods subsection.

3. Construct the correspondence-breaking structural control by replacing the synchronized rPPG identity $i$ with a shuffled identity $\pi(i)$,
\[
S_{ij}^{\mathrm{shuf}}
=\mathcal{S}_j\!\left(z_{ij}^{P},z_{\pi(i)j}^{R}\right).
\]
For similarity statistics, recording specificity requires $S_j^{\mathrm{match}}$ to exceed its shuffled counterpart in the implemented comparison; for distance/discrepancy statistics, the expected direction is reversed. The shuffled control breaks recording identity while retaining the analyzed signal collection.

4. Evaluate separately whether any candidate statistic is intrinsically non-discriminative under the implemented parameterization. For a statistic $q_j$, if
\[
\operatorname{Var}(q_j)\approx0
\]
or its values are effectively concentrated at an admissible boundary, do not interpret that statistic as evidence for or against recording-specific correspondence. In the reported analysis this rule applies to the saturated determinism statistic; recurrence-rate discrepancy remains part of the matched-versus-shuffled structural comparison in Step~3.

5. For the maximal-Lyapunov property, obtain $\lambda_i^{P}$ and $\lambda_i^{R}$ from Algorithm~\ref{alg:lyapunov} and restrict the paired analysis to the common valid index set
\[
\mathcal{I}_{\lambda}=\{i:\lambda_i^{P}\ \text{and}\ \lambda_i^{R}\ \text{are valid}\}.
\]
Compute observed recording-specific correspondence
\[
r_{\lambda}^{\mathrm{obs}}
=\operatorname{corr}\!\left(
\{\lambda_i^{P}\}_{i\in\mathcal{I}_{\lambda}},
\{\lambda_i^{R}\}_{i\in\mathcal{I}_{\lambda}}
\right).
\]

6. Generate the condition-preserving subject-permutation null
\[
r_{\lambda}^{(b)}
=\operatorname{corr}\!\left(
\{\lambda_i^{P}\}_{i\in\mathcal{I}_{\lambda}},
\{\lambda_{\pi_b(i)}^{R}\}_{i\in\mathcal{I}_{\lambda}}
\right),
\qquad b=1,\ldots,B,
\]
and evaluate $r_{\lambda}^{\mathrm{obs}}$ relative to $\{r_{\lambda}^{(b)}\}_{b=1}^{B}$.

7. Report recoverability property by property as the joint evidence from endpoint loss, matched-versus-shuffled structural comparison, recurrence-statistic informativeness, and recording-specific dynamical correspondence. Marginal similarity is treated as population-level plausibility, not by itself as synchronized-recording preservation.

\textbf{Output:} a property-specific recoverability profile
\[
\mathcal{R}=\{\mathcal{R}_{HR},\mathcal{R}_{ACF},\mathcal{R}_{PSD},\mathcal{R}_{REC},\mathcal{R}_{\lambda}\}.
\]
\par\smallskip\noindent\rule{\linewidth}{0.4pt}\par\medskip

\refstepcounter{manuscriptalgorithm}\label{alg:lyapunov}
\textbf{Algorithm~\themanuscriptalgorithm{} \textbar{} Study-specific maximal-Lyapunov estimation and paired discrepancy}

\noindent\textit{Study-specific procedure for estimating $\hat{\lambda}_{\max}$ from each modality under identical validity rules and forming the matched PPG--rPPG dynamical discrepancy.}

\par\smallskip\noindent\rule{\linewidth}{0.4pt}\par\smallskip
\textbf{Input:} one processed waveform $x(t)$ from either contact PPG or camera-derived rPPG, with the preprocessing, analysis interval, SQI gate, and numerical constants specified in the Methods.

1. Let $\tilde{x}(t)=\mathcal{P}[x(t)]$ denote the implemented preprocessing operator, including resampling and filtering. Apply the prespecified spectral-quality gate
\[
Q(\tilde{x})\ge q_0,
\]
where $Q(\cdot)$ and threshold $q_0$ are the study-specific quality criterion defined by the implemented estimator and described in Methods. Waveforms failing this gate do not enter the Lyapunov estimate.

2. Estimate the delay from the average-mutual-information sequence,
\[
\tau=\mathcal{T}\!\left(AMI_{\tilde{x}}(1),\ldots,AMI_{\tilde{x}}(L)\right),
\]
using the implemented first-local-minimum rule and its prespecified fallback.

3. Reconstruct the delay-coordinate trajectory in $\mathbb{R}^{m}$,
\[
\mathbf{X}_{t}
=\big[\tilde{x}_{t},\tilde{x}_{t-\tau},\ldots,
\tilde{x}_{t-(m-1)\tau}\big]^{\!T},
\]
with the embedding dimension $m$ specified in the Methods.

4. Let $T_W$ denote the temporal-exclusion duration and let
\[
W_s=\left\lceil f_sT_W\right\rceil
\]
be the corresponding exclusion width in samples at analysis sampling frequency $f_s$. For each admissible state $\mathbf{X}_{t}$, define its temporally admissible nearest neighbor
\[
j(t)=\arg\min_{s:\,|s-t|>W_s}\left\|\mathbf{X}_{t}-\mathbf{X}_{s}\right\|_2.
\]
In the reported implementation, $T_W=100$ ms after resampling to the common analysis time base.

5. For forward step $k$, compute the ensemble mean log-divergence
\[
\bar d(k)
=\frac{1}{|\mathcal{A}_k|}
\sum_{t\in\mathcal{A}_k}
\log\left\|
\mathbf{X}_{t+k}-\mathbf{X}_{j(t)+k}
\right\|_2,
\]
where $\mathcal{A}_k$ contains neighbor pairs for which both trajectories remain defined at step $k$.

6. Express forward divergence time as
\[
t_k=\frac{k}{f_s},
\]
and estimate the study-specific maximal-divergence slope by the implemented linear fit
\[
\hat\lambda_{\max}
=\operatorname{slope}\{(t_k,\bar d(k))\}_{k\in\mathcal{K}},
\]
where $\mathcal{K}$ is the fitting region used in the reported analysis. Apply the same operational validity rule used to generate the reported results; failed estimates remain missing rather than being imputed.

7. Apply Steps 1--6 independently to $x_i^{P}$ and $x_i^{R}$. For recording $i$, form the paired dynamical discrepancy only when both estimates are valid,
\[
D_{\lambda,i}
=\left|\hat\lambda_{\max,i}^{P}-\hat\lambda_{\max,i}^{R}\right|.
\]
Recording-specific correspondence across the common valid set is then evaluated by Algorithm~\ref{alg:recoverability}.

\textbf{Output:} $\hat\lambda_{\max}$ with validity status for each modality and the paired quantity $D_{\lambda}$ where defined.
\par\smallskip\noindent\rule{\linewidth}{0.4pt}\par\medskip

\refstepcounter{manuscriptalgorithm}\label{alg:heterogeneous}
\textbf{Algorithm~\themanuscriptalgorithm{} \textbar{} Mathematical model of recoverability heterogeneity}

\noindent\textit{Subject-aware mixed-effects framework for quantifying how endpoint and dynamical discrepancies vary across measured observation conditions and for assessing RGB-SNR coefficient sensitivity.}

\par\smallskip\noindent\rule{\linewidth}{0.4pt}\par\smallskip
\textbf{Input:} recording-level discrepancies $D_{HR,i}$ and, where valid, $D_{\lambda,i}$; Fitzpatrick group $F_i$; lighting $L_i$; motion $M_i$; subject $s(i)$; channel-wise SNR indices.

1. Define the two discrepancy variables
\[
D_{HR,i}=\left|HR_i^{R}-HR_i^{P}\right|,
\qquad
D_{\lambda,i}=\left|\lambda_i^{R}-\lambda_i^{P}\right|.
\]
Use the Fitzpatrick-consistent population for models containing $F_i$ and the common valid Lyapunov set for $D_{\lambda}$.

2. Represent repeated measurements by a model-specific subject random intercept $b_{s(i)}$, with
\[
b_s\sim\mathcal{N}(0,\sigma_b^2).
\]
For endpoint loss, fit
\[
D_{HR,i}
=\beta_0+\boldsymbol{\beta}_{F}^{T}F_i
+\boldsymbol{\beta}_{L}^{T}L_i
+\boldsymbol{\beta}_{M}^{T}M_i
+b_{s(i)}+\varepsilon_i,
\]
and for dynamical loss fit the analogous model
\[
D_{\lambda,i}
=\gamma_0+\boldsymbol{\gamma}_{F}^{T}F_i
+\boldsymbol{\gamma}_{L}^{T}L_i
+\boldsymbol{\gamma}_{M}^{T}M_i
+b_{s(i)}+\eta_i.
\]
The random-intercept variance is estimated separately for each fitted model.

3. For the prespecified secondary interaction analyses, augment one condition family at a time. For $D_i\in\{D_{HR,i},D_{\lambda,i}\}$, the Fitzpatrick-by-motion model is
\[
D_i=\alpha_0
+\boldsymbol{\alpha}_{F}^{T}F_i
+\boldsymbol{\alpha}_{L}^{T}L_i
+\boldsymbol{\alpha}_{M}^{T}M_i
+\boldsymbol{\alpha}_{FM}^{T}(F_i\otimes M_i)
+b_{s(i)}+e_i,
\]
whereas the Fitzpatrick-by-lighting model replaces the interaction term by
\[
\boldsymbol{\alpha}_{FL}^{T}(F_i\otimes L_i).
\]
Test the full interaction coefficient block jointly rather than interpreting selected cells post hoc.

4. Define the descriptive aggregate channel-quality index
\[
SNR_{RGB,i}
=\frac{SNR_{R,i}+SNR_{G,i}+SNR_{B,i}}{3}.
\]
On exactly the endpoint-model population, fit the SNR-adjusted sensitivity model
\[
D_{HR,i}
=\theta_0+\boldsymbol{\theta}_{F}^{T}F_i
+\boldsymbol{\theta}_{L}^{T}L_i
+\boldsymbol{\theta}_{M}^{T}M_i
+\theta_S SNR_{RGB,i}
+b_{s(i)}+\xi_i.
\]

5. Let $\beta_{VI,1}$ denote the Fitzpatrick VI-versus-III coefficient before SNR adjustment and $\beta_{VI,1b}$ the corresponding coefficient after SNR adjustment. Quantify descriptive attenuation as
\[
A_{VI}=100\left(
\frac{\beta_{VI,1}-\beta_{VI,1b}}{\beta_{VI,1}}
\right)\%.
\]
Interpret $A_{VI}$ as coefficient sensitivity to aggregate SNR adjustment, not as a causal mediation quantity.

\textbf{Output:} fixed-effect contrasts for endpoint and dynamical loss, joint interaction tests, and the SNR-adjusted coefficient-sensitivity measure $A_{VI}$.
\par\smallskip\noindent\rule{\linewidth}{0.4pt}\par\medskip

\refstepcounter{manuscriptalgorithm}\label{alg:conditionaware}
\textbf{Algorithm~\themanuscriptalgorithm{} \textbar{} Mathematical formulation of condition-aware endpoint-discrepancy reduction}

\noindent\textit{Subject-disjoint prediction procedure for testing whether incrementally added observation context reduces held-out PPG--rPPG endpoint discrepancy, with MAE as the primary performance measure.}

\par\smallskip\noindent\rule{\linewidth}{0.4pt}\par\smallskip
\textbf{Input:} contact-PPG target $Y_i=HR_i^{P}$, camera-derived HR $H_i=HR_i^{R}$, observation variables $(M_i,L_i,F_i,S_i)$, and subject identifier $s(i)$, with $S_i=SNR_{RGB,i}$.

1. Define the nested information sets
\[
\mathcal{X}_0=\{H\},\qquad
\mathcal{X}_1=\{H,M\},\qquad
\mathcal{X}_2=\{H,M,L\},
\]
\[
\mathcal{X}_3=\{H,M,L,F\},\qquad
\mathcal{X}_4=\{H,M,L,F,S\}.
\]
Thus each successive model asks whether an additional observation variable contains predictive information about contact-PPG HR beyond the preceding set.

2. Partition the subject set into five disjoint folds $\mathcal{G}_1,\ldots,\mathcal{G}_5$. For fold $g$, define
\[
\mathcal{D}_{test}^{(g)}=\{i:s(i)\in\mathcal{G}_g\},
\qquad
\mathcal{D}_{train}^{(g)}=\{i:s(i)\notin\mathcal{G}_g\},
\]
so that
\[
\{s(i):i\in\mathcal{D}_{train}^{(g)}\}
\cap
\{s(i):i\in\mathcal{D}_{test}^{(g)}\}=\varnothing.
\]

3. For learner family $\ell$ and information set $\mathcal{X}_k$, estimate a fold-specific prediction function using training subjects only,
\[
\hat f_{k,\ell}^{(g)}
=\mathcal{A}_{\ell}
\left(\mathcal{D}_{train}^{(g)},\mathcal{X}_k,Y\right),
\]
where $\mathcal{A}_{\ell}$ denotes the implemented linear, ridge, or random-forest learner with the preprocessing specified in Methods.

4. Generate exactly one held-out prediction for each recording,
\[
\hat Y_{i,k,\ell}
=\hat f_{k,\ell}^{(g)}(X_{i,k}),
\qquad i\in\mathcal{D}_{test}^{(g)},
\]
and concatenate predictions over $g=1,\ldots,5$.

5. Evaluate each $(k,\ell)$ using the pooled out-of-fold errors
\[
e_{i,k,\ell}=Y_i-\hat Y_{i,k,\ell},
\]
Let $\mathcal{D}_{\mathrm{eval}}$ denote the complete out-of-fold evaluation population and $N_{\mathrm{eval}}=|\mathcal{D}_{\mathrm{eval}}|$. Then
\[
MAE_{k,\ell}=\frac{1}{N_{\mathrm{eval}}}\sum_{i\in\mathcal{D}_{\mathrm{eval}}}|e_{i,k,\ell}|,
\qquad
RMSE_{k,\ell}=\sqrt{\frac{1}{N_{\mathrm{eval}}}\sum_{i\in\mathcal{D}_{\mathrm{eval}}}e_{i,k,\ell}^{2}},
\]
along with the reported median absolute error, Pearson correlation, and $R^2$.

6. Within each learner family, treat $\mathcal{X}_0$ as the learned camera-HR calibration baseline and define the incremental endpoint benefit
\[
\Delta MAE_{k,\ell}
=MAE_{k,\ell}-MAE_{0,\ell},
\qquad k=1,\ldots,4.
\]
Hence $\Delta MAE_{k,\ell}<0$ denotes lower subject-held-out error after adding observation context.

7. Let $\mathcal{B}^{(b)}$ be a bootstrap resample of subjects. Using the fixed out-of-fold predictions, recompute
\[
\Delta MAE_{k,\ell}^{(b)}
=MAE_{k,\ell}^{(b)}-MAE_{0,\ell}^{(b)},
\qquad b=1,\ldots,B.
\]
Form the reported confidence interval from the empirical distribution of $\{\Delta MAE_{k,\ell}^{(b)}\}_{b=1}^{B}$.

8. Declare evidence of incremental held-out MAE reduction only when the complete confidence interval for $\Delta MAE_{k,\ell}$ lies below zero. This criterion concerns endpoint prediction only and is not interpreted as waveform reconstruction or PPG--rPPG equivalence.

\textbf{Output:} subject-held-out performance for $\mathcal{X}_0$--$\mathcal{X}_4$ and uncertainty-calibrated incremental endpoint benefit $\Delta MAE_{k,\ell}$.
\par\smallskip\noindent\rule{\linewidth}{0.4pt}\par\medskip

\section{Results}\label{results}

\subsection{Endpoint recovery under the fixed observation pathway}\label{dataset-accounting-and-frozen-chrom-benchmark}

The MMPD design comprises 33 participants with 20 recording indices (0--19), yielding 660 expected subject-recording combinations. Five files were unavailable, leaving 655 recordings for the fixed-pathway benchmark. The observation pipeline processed all 655 available recordings without technical failure. The dynamics pipeline retained all 33 subjects and assigned an explicit success or failure status to every expected combination; Subject 24 was excluded only from Fitzpatrick-stratified/model-based analyses because its skin-tone metadata were internally inconsistent, not from the dynamics computation. As summarized in Table~\ref{tab:chrom}, HR recovery
remained weak: MAE was 15.26 beats/min, RMSE 20.81 beats/min, MAPE
17.97\%, and Pearson r=0.0801. Mean signed error was -6.80 beats/min.
Reference HR averaged 83.06 beats/min (SD 15.88), compared with 76.27
beats/min (SD 12.96) for camera-derived rPPG. As an implementation check, these results were compared with the published CHROM benchmark on MMPD reported through the rPPG-Toolbox framework: MAE 13.66 $\pm$ 0.50 bpm, RMSE 18.76 $\pm$ 23.82 bpm, MAPE 16.00 $\pm$ 0.57\%, and Pearson $r=0.08 \pm 0.04$ \cite{tang2023mmpd,liu2023rppgtoolbox}. The present Pearson correlation ($r=0.0801$) therefore closely reproduces the published correlation, while the present MAE (15.26 bpm) is modestly higher than the published 13.66-bpm benchmark. This agreement in correlation and broadly comparable absolute-error regime supported use of the same camera-to-rPPG implementation for the downstream recoverability analyses.

\begin{longtable}{@{}
 >{\raggedright\arraybackslash}p{(\columnwidth - 2\tabcolsep) * \real{0.5000}}
 >{\raggedright\arraybackslash}p{(\columnwidth - 2\tabcolsep) * \real{0.5000}}@{}}

\multicolumn{2}{@{}p{\linewidth}@{}}{\textbf{Table~\thetable{} \textbar{} Fixed camera-rPPG observation benchmark on the 655 available MMPD recordings.}\par\vspace{2pt}\textit{Heart rate from the fixed camera-rPPG observation pathway is compared with reference HR derived from synchronized contact PPG. MAE, RMSE, and MAPE quantify absolute endpoint error; Pearson $r$ quantifies recording-level linear correspondence; mean signed error indicates systematic underestimation when negative. The benchmark is used as an implementation and endpoint-recoverability check rather than as an optimization result.}} \\[4pt]
\toprule\noalign{}
\begin{minipage}[b]{\linewidth}\raggedright
\textbf{Metric}
\end{minipage} & \begin{minipage}[b]{\linewidth}\raggedright
\textbf{Fixed camera-rPPG observation benchmark}
\end{minipage} \\
\midrule\noalign{}
\endhead
\bottomrule\noalign{}
\endlastfoot
\label{tab:chrom}Available recordings & 655 \\
MAE & 15.26 bpm \\
RMSE & 20.81 bpm \\
MAPE & 17.97\% \\
Pearson r & 0.0801 \\
Mean SNR & -6.56 dB \\
Mean signed error & -6.80 bpm \\
GT HR mean (SD) & 83.06 (15.88) bpm \\
rPPG HR mean (SD) & 76.27 (12.96) bpm \\
\end{longtable}

\subsection{Conventional structure shows selective recording-specific correspondence}\label{conventional-signal-structure-shows-selective-recording-specificity}

Table~\ref{tab:structural} summarizes the matched-versus-shuffled validation used to determine whether
apparent structural similarity was recording-specific. ACF similarity
was 0.2863 for matched pairs and 0.2440 after shuffling, a difference of
+0.0422. Cardiac-band PSD distance was 0.0515 for matched pairs and
0.0524 for shuffled pairs (difference -0.0009), while recurrence-rate
discrepancy was 0.0246 and 0.0263, respectively (difference -0.0017).
Thus, temporal autocorrelation retained modest recording specificity,
whereas the spectral and recurrence-rate measures provided little
separation from the shuffled control. Determinism was identically 1.0
and was excluded because it was non-discriminative.

\begin{longtable}{@{}
 >{\raggedright\arraybackslash}p{(\columnwidth - 6\tabcolsep) * \real{0.2500}}
 >{\raggedright\arraybackslash}p{(\columnwidth - 6\tabcolsep) * \real{0.2500}}
 >{\raggedright\arraybackslash}p{(\columnwidth - 6\tabcolsep) * \real{0.2500}}
 >{\raggedright\arraybackslash}p{(\columnwidth - 6\tabcolsep) * \real{0.2500}}@{}}

\multicolumn{4}{@{}p{\linewidth}@{}}{\textbf{Table~\thetable{} \textbar{} Matched-versus-shuffled structural recoverability analysis.}\par\vspace{2pt}\textit{Matched PPG--rPPG property correspondence is compared with a shuffled-pair null distribution to distinguish recording-specific preservation from cross-recording similarity. Larger matched-versus-shuffled separation indicates stronger evidence of recording-specific structural correspondence.}} \\[4pt]
\toprule\noalign{}
\begin{minipage}[b]{\linewidth}\raggedright
\textbf{Metric}
\end{minipage} & \begin{minipage}[b]{\linewidth}\raggedright
\textbf{Matched}
\end{minipage} & \begin{minipage}[b]{\linewidth}\raggedright
\textbf{Shuffled}
\end{minipage} & \begin{minipage}[b]{\linewidth}\raggedright
\textbf{Matched-shuffled interpretation}
\end{minipage} \\
\midrule\noalign{}
\endhead
\bottomrule\noalign{}
\endlastfoot
\label{tab:structural}ACF similarity & 0.2863 & 0.2440 & +0.0422; modest recording
specificity \\
PSD distance & 0.0515 & 0.0524 & -0.0009; marginal \\
Recurrence-rate difference & 0.0246 & 0.0263 & -0.0017; marginal \\
Determinism & 1.000 & 1.000 & Non-discriminative; excluded \\
\end{longtable}

\subsection{Recording-specific maximal-Lyapunov correspondence is absent}\label{recording-specific-maximal-lyapunov-correspondence-is-absent}

The study-specific \ensuremath{\lambda}max estimator was applied identically to PPG and
camera-derived rPPG. The primary analysis used the exact intersection of
recordings with valid estimates in both modalities. As reported in Table~\ref{tab:lyapunov}, across 645 matched
recordings, recording-specific correspondence was essentially absent
(r\ensuremath{\lambda}=0.0231, 95\% CI -0.0542 to 0.1001; p=0.5579). A 10,000-iteration
condition-preserving subject permutation test produced p=0.5584,
providing no evidence that the observed correspondence exceeded that
expected after breaking subject identity.

\begin{longtable}{@{}
 >{\raggedright\arraybackslash}p{(\columnwidth - 8\tabcolsep) * \real{0.2000}}
 >{\raggedright\arraybackslash}p{(\columnwidth - 8\tabcolsep) * \real{0.2000}}
 >{\raggedright\arraybackslash}p{(\columnwidth - 8\tabcolsep) * \real{0.2000}}
 >{\raggedright\arraybackslash}p{(\columnwidth - 8\tabcolsep) * \real{0.2000}}
 >{\raggedright\arraybackslash}p{(\columnwidth - 8\tabcolsep) * \real{0.2000}}@{}}

\multicolumn{5}{@{}p{\linewidth}@{}}{\textbf{Table~\thetable{} \textbar{} Recording-specific maximal-Lyapunov correspondence test.}\par\vspace{2pt}\textit{Pearson correlation quantifies matched recording-level correspondence between valid contact-PPG and camera-rPPG maximal-Lyapunov estimates. The confidence interval and permutation test assess whether the observed correspondence differs detectably from zero and from correspondence expected under disrupted pairing.}} \\[4pt]
\toprule\noalign{}
\begin{minipage}[b]{\linewidth}\raggedright
\textbf{N matched}
\end{minipage} & \begin{minipage}[b]{\linewidth}\raggedright
\textbf{r\_lambda}
\end{minipage} & \begin{minipage}[b]{\linewidth}\raggedright
\textbf{95\% CI}
\end{minipage} & \begin{minipage}[b]{\linewidth}\raggedright
\textbf{Pearson p}
\end{minipage} & \begin{minipage}[b]{\linewidth}\raggedright
\textbf{Permutation p}
\end{minipage} \\
\midrule\noalign{}
\endhead
\bottomrule\noalign{}
\endlastfoot
\label{tab:lyapunov}645 & 0.0231 & {[}-0.0542, 0.1001{]} & 0.5579 & 0.5584 \\
\end{longtable}

Marginal similarity did not alter this conclusion. Mean \ensuremath{\lambda}max was
approximately 2.063 for PPG and 2.256 for rPPG, but the corresponding
SDs were 0.9795 and 0.2809. Thus, rPPG retained only about 29\% of the
PPG between-recording variability in this descriptor. The combination of
near-zero matched correlation and strong variance compression indicates
absence of recording-specific dynamical correspondence despite overlap in the
marginal value ranges. The matched and marginal maximal-Lyapunov results are visualized in Supplementary Figure~\ref{fig:s4}.

\subsection{Observation-associated discrepancy is property-specific}

\subsubsection{Endpoint discrepancy across Fitzpatrick skin type}
\label{endpoint-recoverability-is-heterogeneous-across-fitzpatrick-skin-type}

After removal of one duplicate match key, 645 unique matched records
remained. Subject 24 was excluded only from Fitzpatrick analyses because
its raw Fitzpatrick-group metadata were inconsistent, leaving 625 recordings
from 32 subjects. Table~\ref{tab:fitz-unadjusted} shows that mean absolute HR discrepancy (DHR) increased from
11.43 beats/min for Fitzpatrick III to 20.64 beats/min for Fitzpatrick
VI. By contrast, mean absolute \ensuremath{\lambda}max discrepancy (D\ensuremath{\lambda}) remained
approximately flat across Fitzpatrick groups.

\begin{table}[!htbp]
\centering
\refstepcounter{table}\label{tab:fitz-unadjusted}
\noindent\parbox{\linewidth}{\textbf{Table~\thetable{} \textbar{} Unadjusted Fitzpatrick-stratified endpoint and dynamical discrepancies for the 625-record, 32-subject analysis set.}\par\vspace{2pt}\textit{$D_{HR}$ is the absolute contact-PPG versus camera-rPPG heart-rate discrepancy in beats/min; $D_{\lambda}$ is the absolute discrepancy in maximal Lyapunov exponent. Subject 24 is excluded from this stratified analysis because of inconsistent Fitzpatrick metadata. These summaries are descriptive and do not adjust for repeated measures, lighting, or motion/activity.}}\par\vspace{4pt}
\resizebox{0.94\linewidth}{!}{%
\begin{tabular}{@{}lrrrrrr@{}}
\toprule
\textbf{Fitzpatrick} & \textbf{Subjects} & \textbf{N} &
\textbf{Mean $D_{HR}$} & \textbf{Median $D_{HR}$} &
\textbf{Mean $D_{\lambda}$} & \textbf{Median $D_{\lambda}$} \\
\midrule
III & 16 & 315 & 11.43 & 8.79 & 0.8187 & 0.6238 \\
IV & 4 & 75 & 18.01 & 15.82 & 0.8082 & 0.7274 \\
V & 6 & 119 & 18.29 & 16.70 & 0.8016 & 0.6058 \\
VI & 6 & 116 & 20.64 & 18.46 & 0.8246 & 0.6801 \\
\bottomrule
\end{tabular}%
}
\end{table}

\subsubsection{Subject-aware adjusted models confirm a Fitzpatrick endpoint
gradient}\label{subject-aware-adjusted-models-confirm-a-fitzpatrick-endpoint-gradient}

In Model 1, DHR was modeled as a function of Fitzpatrick group,
lighting, and motion/activity, with a subject-specific random intercept.
The adjusted Model 1 contrasts are reported in Table~\ref{tab:model1}. Relative to Fitzpatrick III, the adjusted difference was +6.37 beats/min
for IV (95\% CI -0.12 to 12.87), +6.72 for V (95\% CI 1.19 to 12.25),
and +9.32 for VI (95\% CI 3.78 to 14.86). The increasing endpoint
discrepancy therefore persisted after adjustment for measured lighting
and motion conditions. A coefficient-only display of the Fitzpatrick contrasts is provided in Supplementary Figure~\ref{fig:s1}.

\begin{longtable}{@{}
 >{\raggedright\arraybackslash}p{(\columnwidth - 6\tabcolsep) * \real{0.2500}}
 >{\raggedright\arraybackslash}p{(\columnwidth - 6\tabcolsep) * \real{0.2500}}
 >{\raggedright\arraybackslash}p{(\columnwidth - 6\tabcolsep) * \real{0.2500}}
 >{\raggedright\arraybackslash}p{(\columnwidth - 6\tabcolsep) * \real{0.2500}}@{}}

\multicolumn{4}{@{}p{\linewidth}@{}}{\textbf{Table~\thetable{} \textbar{} Subject-aware Model 1 Fitzpatrick contrasts for endpoint discrepancy.}\par\vspace{2pt}\textit{Estimates are adjusted differences in absolute PPG--rPPG HR discrepancy ($D_{HR}$), in beats/min, relative to Fitzpatrick III after adjustment for lighting and motion/activity, with a subject-specific random intercept. Positive estimates indicate larger endpoint discrepancy than the Fitzpatrick III reference group.}} \\[4pt]
\toprule\noalign{}
\begin{minipage}[b]{\linewidth}\raggedright
\textbf{Contrast}
\end{minipage} & \begin{minipage}[b]{\linewidth}\raggedright
\textbf{Estimate (bpm)}
\end{minipage} & \begin{minipage}[b]{\linewidth}\raggedright
\textbf{95\% CI}
\end{minipage} & \begin{minipage}[b]{\linewidth}\raggedright
\textbf{p}
\end{minipage} \\
\midrule\noalign{}
\endhead
\bottomrule\noalign{}
\endlastfoot
\label{tab:model1}IV vs III & +6.37 & {[}-0.12, 12.87{]} & 0.051 \\
V vs III & +6.72 & {[}1.19, 12.25{]} & 0.017 \\
VI vs III & +9.32 & {[}3.78, 14.86{]} & 0.001 \\
\end{longtable}

Lighting effects were smaller than the largest Fitzpatrick and motion
effects. Relative to incandescent lighting, LED-low was associated with
+3.06 beats/min (95\% CI 0.47 to 5.66) and natural lighting with +2.93
beats/min (95\% CI 0.31 to 5.55). Relative to rotation, walking was
associated with +9.69 beats/min (95\% CI 6.77 to 12.60), and
stationary-after-exercise with +9.05 beats/min (95\% CI 6.13 to 11.98).

\begin{figure}[!htbp]
\centering
\includegraphics[width=\textwidth]{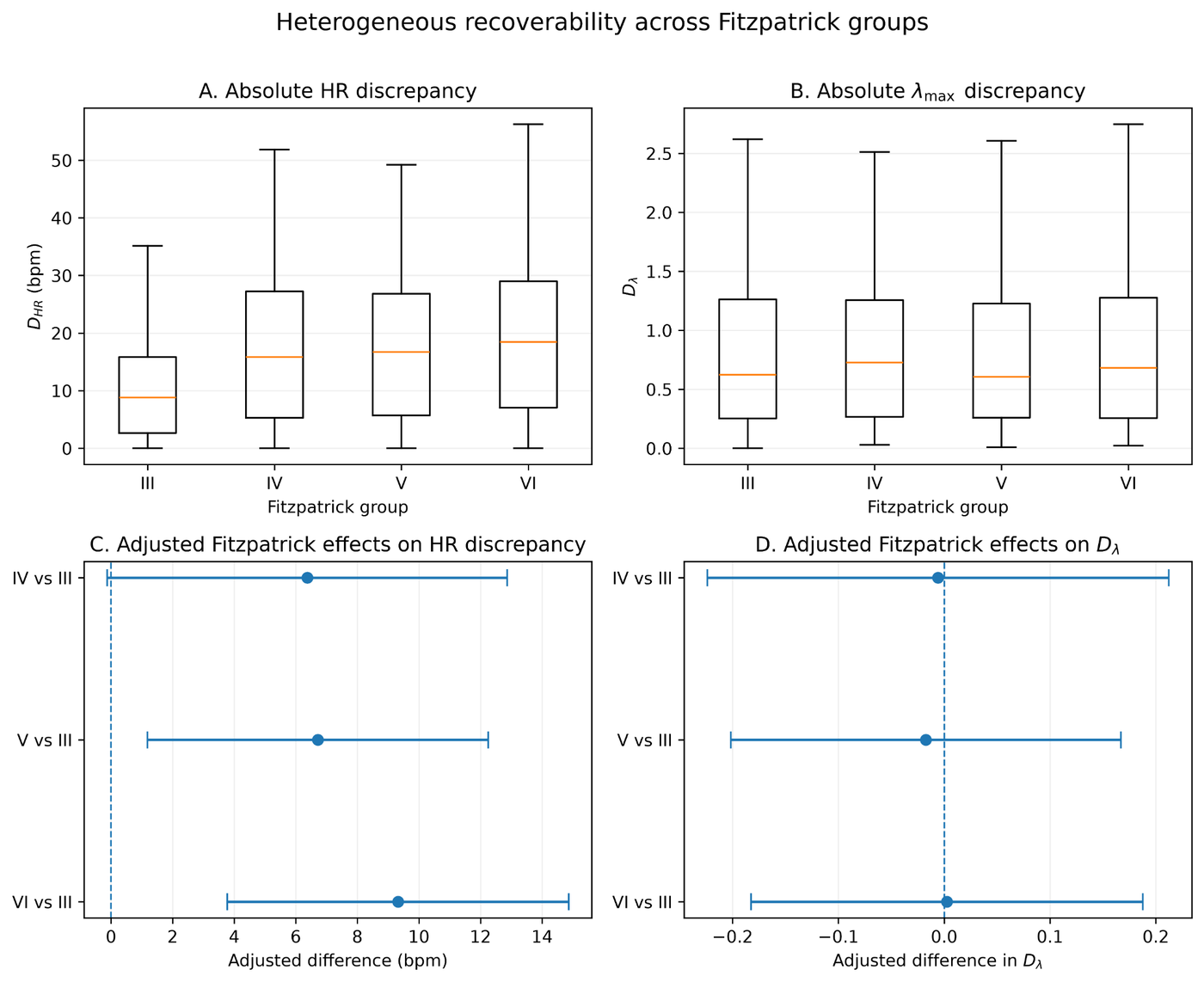}
\caption{\textbf{Heterogeneous recoverability of endpoint and dynamical quantities across Fitzpatrick groups.} \textit{(A) Distribution of absolute PPG--rPPG heart-rate discrepancy, $D_{HR}$, across Fitzpatrick groups. (B) Distribution of absolute maximal-Lyapunov discrepancy, $D_{\lambda}$. (C) Subject-aware adjusted Fitzpatrick contrasts for $D_{HR}$ relative to Fitzpatrick III after adjustment for lighting and motion/activity. (D) Corresponding adjusted contrasts for $D_{\lambda}$. The endpoint discrepancy increases from Fitzpatrick III toward higher-numbered Fitzpatrick groups, whereas the tested dynamical discrepancy does not exhibit the same gradient. Error bars in C--D denote 95\% confidence intervals; contrasts are relative to Fitzpatrick III. Panels A--B are unadjusted distributions, whereas C--D are estimates from subject-aware mixed-effects models adjusted for lighting and motion/activity. Fitzpatrick analyses include 625 recordings from 32 subjects after exclusion of the subject with inconsistent Fitzpatrick-group metadata. The contrast between the endpoint and dynamical panels is interpreted as property-specific heterogeneity, not as evidence that nonlinear dynamics are preserved.}}
\label{fig:heterogeneous}
\end{figure}

\clearpage
\subsubsection{The dynamical discrepancy does not show the same Fitzpatrick
gradient}\label{the-dynamical-discrepancy-does-not-show-the-same-fitzpatrick-gradient}

Model 2 applied the same mixed-effects structure to D\ensuremath{\lambda}. The adjusted Model 2 contrasts are reported in Table~\ref{tab:model2}. Relative to
Fitzpatrick III, coefficients were -0.006 for IV (95\% CI -0.224 to
0.212), -0.017 for V (95\% CI -0.202 to 0.167), and +0.003 for VI (95\%
CI -0.183 to 0.188); all p-values exceeded 0.80. The absence of a corresponding Fitzpatrick gradient is visualized in Figure~\ref{fig:heterogeneous}B,D. This result should not
be interpreted as preservation of nonlinear dynamics, because
recording-specific \ensuremath{\lambda}max correspondence was already absent overall.
Rather, the magnitude of the PPG-rPPG \ensuremath{\lambda}max discrepancy did not exhibit
an additional Fitzpatrick-associated gradient.

\begin{longtable}{@{}
 >{\raggedright\arraybackslash}p{(\columnwidth - 6\tabcolsep) * \real{0.2500}}
 >{\raggedright\arraybackslash}p{(\columnwidth - 6\tabcolsep) * \real{0.2500}}
 >{\raggedright\arraybackslash}p{(\columnwidth - 6\tabcolsep) * \real{0.2500}}
 >{\raggedright\arraybackslash}p{(\columnwidth - 6\tabcolsep) * \real{0.2500}}@{}}

\multicolumn{4}{@{}p{\linewidth}@{}}{\textbf{Table~\thetable{} \textbar{} Subject-aware Model 2 Fitzpatrick contrasts for dynamical discrepancy.}\par\vspace{2pt}\textit{Estimates are adjusted differences in absolute PPG--rPPG maximal-Lyapunov discrepancy ($D_{\lambda}$) relative to Fitzpatrick III after adjustment for lighting and motion/activity, with a subject-specific random intercept. Confidence intervals and $p$-values quantify uncertainty in the adjusted contrasts.}} \\[4pt]
\toprule\noalign{}
\begin{minipage}[b]{\linewidth}\raggedright
\textbf{Contrast}
\end{minipage} & \begin{minipage}[b]{\linewidth}\raggedright
\textbf{D\_lambda estimate}
\end{minipage} & \begin{minipage}[b]{\linewidth}\raggedright
\textbf{95\% CI}
\end{minipage} & \begin{minipage}[b]{\linewidth}\raggedright
\textbf{p}
\end{minipage} \\
\midrule\noalign{}
\endhead
\bottomrule\noalign{}
\endlastfoot
\label{tab:model2}IV vs III & -0.006 & {[}-0.224, 0.212{]} & 0.955 \\
V vs III & -0.017 & {[}-0.202, 0.167{]} & 0.856 \\
VI vs III & +0.003 & {[}-0.183, 0.188{]} & 0.980 \\
\end{longtable}

\begin{figure}[!htbp]
\centering
\includegraphics[width=\textwidth]{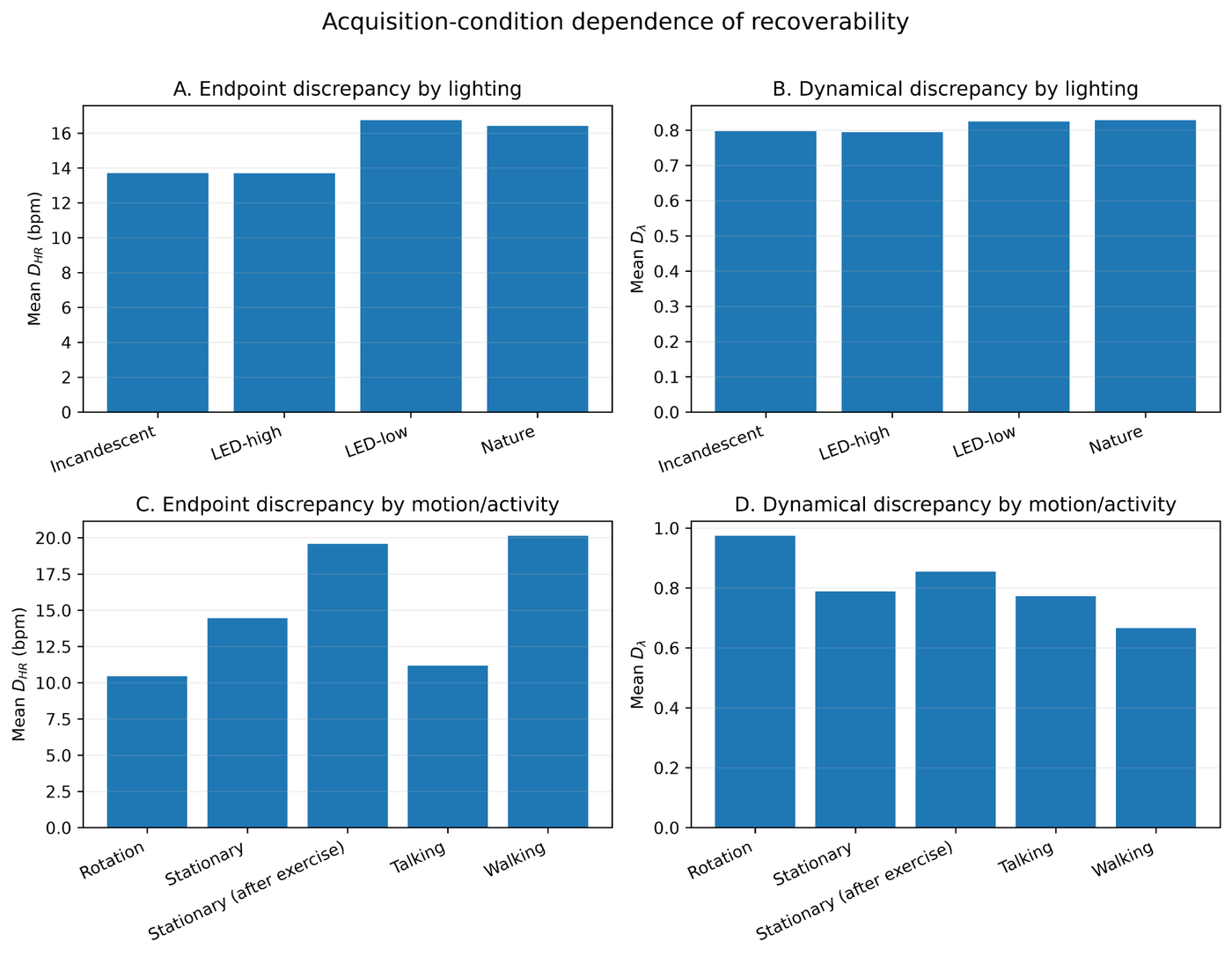}
\caption{\textbf{Acquisition-condition dependence of endpoint and dynamical recoverability.} \textit{Mean $D_{HR}$ and mean $D_{\lambda}$ are shown across lighting conditions (A--B) and motion/activity conditions (C--D). The two discrepancy measures do not vary in parallel across acquisition conditions, motivating property-specific rather than single-metric characterization of PPG-to-rPPG recoverability. Bars summarize observed condition-specific means from the analysis. $D_{HR}$ is expressed in beats/min, whereas $D_{\lambda}$ is dimensionless under the estimator used; their numerical magnitudes are therefore not directly comparable across axes. The panels are intended to compare patterns across conditions, are descriptive, and should not be interpreted as causal effects.}}
\label{fig:acquisition}
\end{figure}

Figure~\ref{fig:acquisition} compares endpoint and dynamical discrepancy across lighting (panels A--B) and motion/activity (panels C--D), showing that the two properties do not vary in parallel across acquisition conditions.

\subsubsection{Fitzpatrick-by-condition interactions are specific to endpoint motion heterogeneity}\label{fitzpatrick-condition-interactions}

The follow-up interaction experiment tested whether the effect of motion/activity or lighting on discrepancy depended on Fitzpatrick group. Cell counts were inspected before inference; all Fitzpatrick-by-motion and Fitzpatrick-by-lighting combinations were represented, although the number of independent participants remained limited in Fitzpatrick IV--VI.

For $D_{HR}$, the joint Fitzpatrick$\times$motion interaction block was significant (Wald $\chi^2=28.789$, df=12, $p=0.004234$), indicating that motion-related changes in endpoint discrepancy were not constant across Fitzpatrick groups. In contrast, the Fitzpatrick$\times$lighting block was not significant (Wald $\chi^2=9.970$, df=9, $p=0.352900$); therefore, the data do not support a compounded Fitzpatrick-by-lighting effect on endpoint discrepancy.

The same interaction structure was not detected for $D_{\lambda}$. The Fitzpatrick$\times$motion block yielded Wald $\chi^2=9.468$ (df=12, $p=0.662521$), and the Fitzpatrick$\times$lighting block yielded Wald $\chi^2=8.534$ (df=9, $p=0.481370$). Thus, the significant endpoint interaction did not generalize to the tested dynamical discrepancy, providing a second level of endpoint--dynamical dissociation beyond the additive contrasts. The four joint interaction tests are summarized in Table~\ref{tab:interactions}.

\begingroup
\small
\begin{longtable}{@{}lcrr@{}}

\multicolumn{4}{@{}p{\linewidth}@{}}{\textbf{Table~\thetable{} \textbar{} Joint Wald tests of prespecified Fitzpatrick-by-observation-condition interaction blocks.}\par\vspace{2pt}\textit{A significant block test indicates heterogeneity of the condition effect across Fitzpatrick groups; nonsignificant blocks are not interpreted from individual cell means alone.}} \\[4pt]
\toprule
\textbf{Outcome and interaction block} & \textbf{df} & \textbf{Wald $\chi^2$} & \textbf{p} \\
\midrule
\endhead
\bottomrule
\endlastfoot
\label{tab:interactions}$D_{HR}$: Fitzpatrick $\times$ Motion & 12 & 28.789 & 0.004234 \\
$D_{HR}$: Fitzpatrick $\times$ Lighting & 9 & 9.970 & 0.352900 \\
$D_{\lambda}$: Fitzpatrick $\times$ Motion & 12 & 9.468 & 0.662521 \\
$D_{\lambda}$: Fitzpatrick $\times$ Lighting & 9 & 8.534 & 0.481370 \\
\end{longtable}
\endgroup

\begin{figure}[!htbp]
\centering
\includegraphics[width=\textwidth]{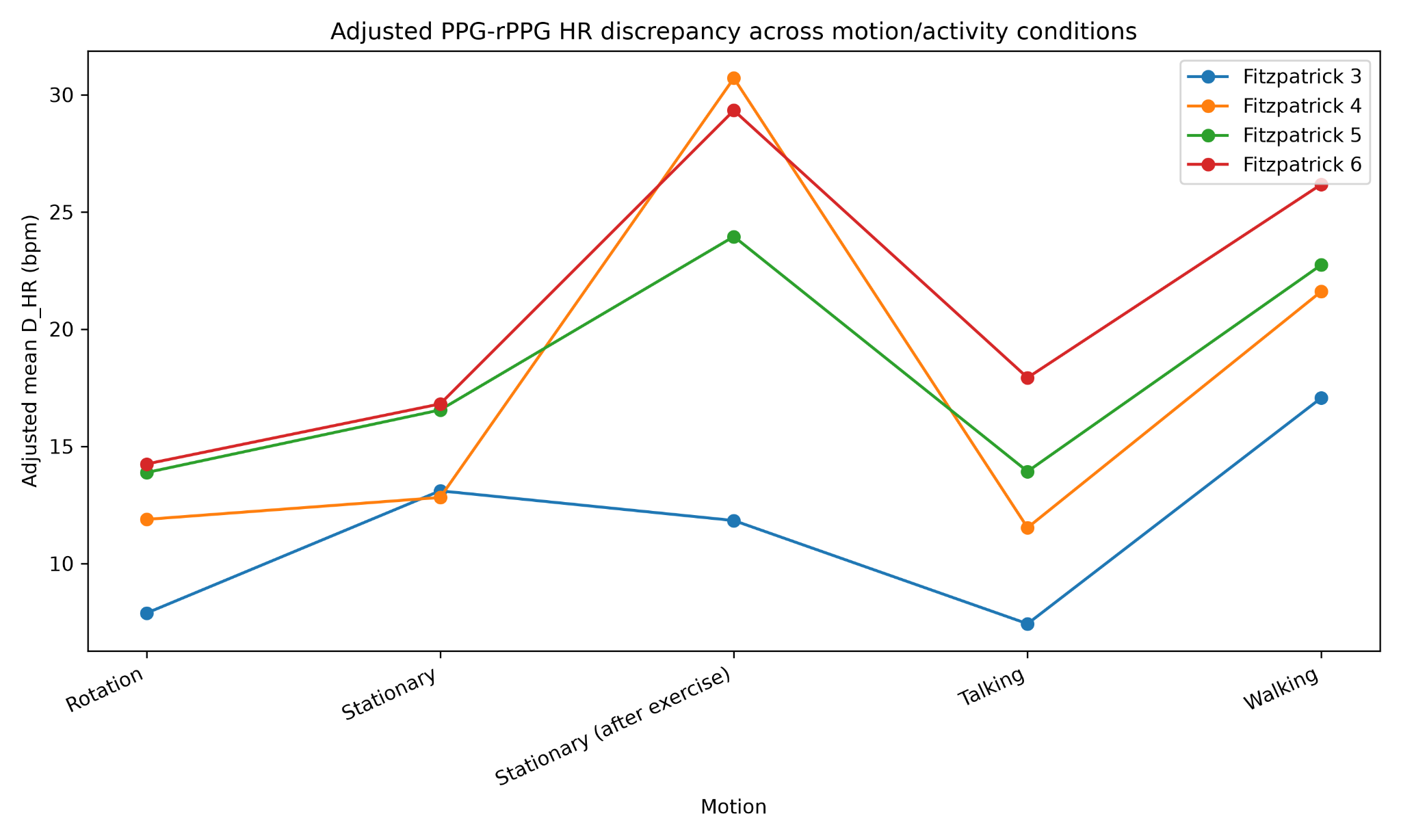}
\caption{\textbf{Adjusted Fitzpatrick-by-motion heterogeneity in endpoint recoverability.} \textit{Population-level adjusted mean absolute PPG--rPPG HR discrepancy ($D_{HR}$) from the subject-aware Fitzpatrick$\times$motion model is shown across motion/activity conditions. The nonparallel profiles visualize the significant joint interaction block (Wald $\chi^2(12)=28.789$, $p=0.004234$). The largest separation is observed in the post-exercise stationary condition, where adjusted discrepancy rises sharply for Fitzpatrick IV--VI relative to III. The figure visualizes population-level model-based adjusted means from the repeated-measures model; inferential interpretation is based on the joint 12-df interaction block test rather than on isolated plotted cell differences. A significant block indicates that the motion/activity pattern differs across Fitzpatrick groups, but does not imply that every motion contrast differs for every group.}}
\label{fig:fitzmotion}
\end{figure}

The resulting adjusted profiles are shown in Figure~\ref{fig:fitzmotion}. The nonparallel motion profiles visualize the significant interaction, with the largest separation occurring in the post-exercise stationary condition.

The significant interaction is therefore not a generic claim that every adverse condition amplifies a Fitzpatrick-associated gap. The corresponding Fitzpatrick$\times$lighting block was nonsignificant despite visible variation among adjusted cell means (Supplementary Figure~\ref{fig:s6}). This contrast is important because it separates an interaction supported by the repeated-measures model from visually suggestive but statistically unsupported subgroup patterns.

\subsubsection{Aggregate RGB SNR does not materially account for the endpoint
gradient}\label{aggregate-rgb-snr-does-not-explain-the-endpoint-gradient}

\begin{figure}[!htbp]
\centering
\includegraphics[width=\textwidth]{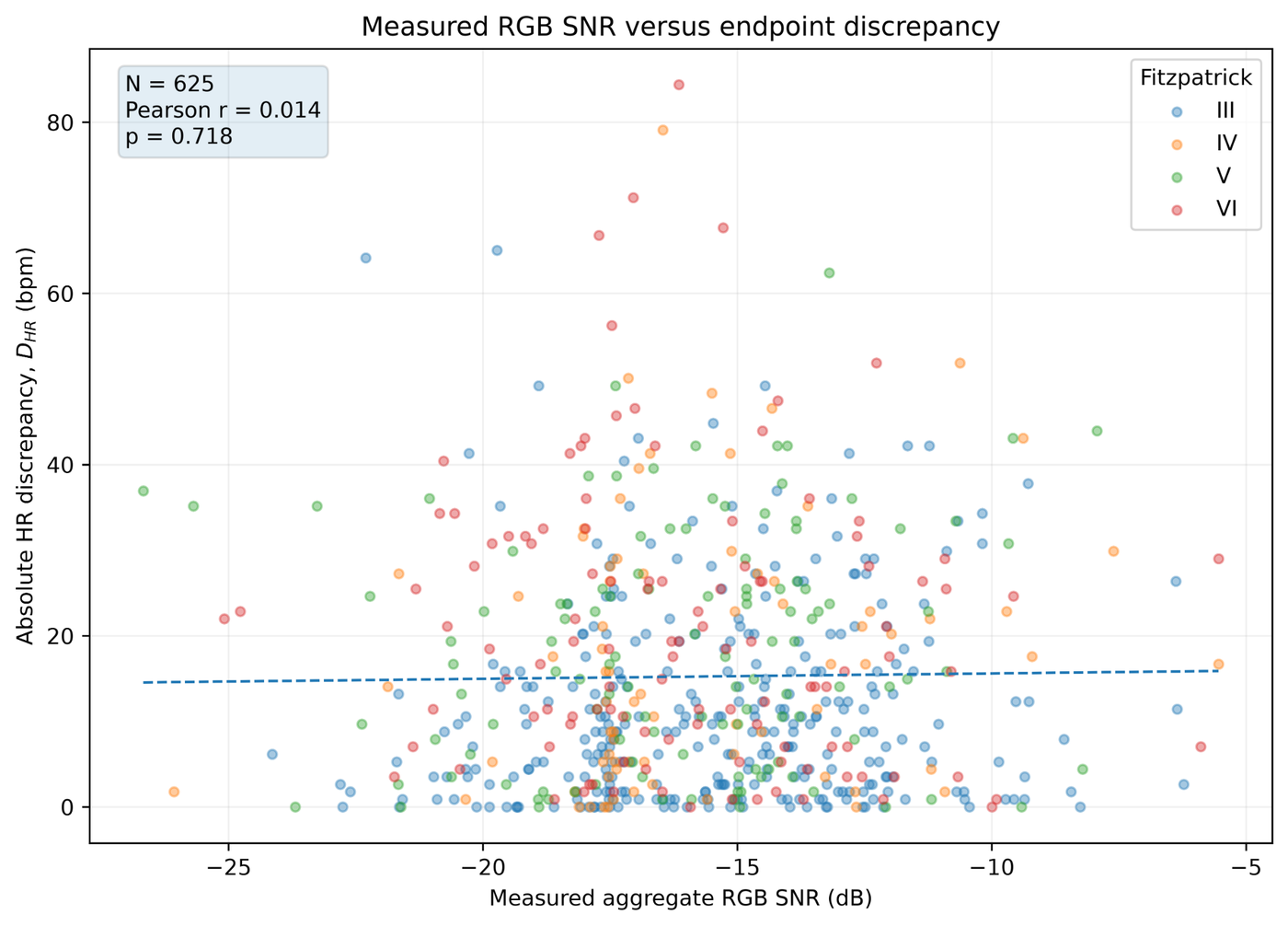}
\caption{\textbf{Measured aggregate RGB SNR versus endpoint discrepancy.} \textit{Recording-level aggregate RGB SNR is plotted against absolute heart-rate discrepancy $D_{HR}$ for the 625-record Fitzpatrick analysis set. The linear association is negligible (Pearson $r=0.014$, $p=0.718$), indicating that the measured aggregate SNR metric does not materially account for recording-level variation in endpoint discrepancy. Points are individual recordings and are identified by Fitzpatrick group for visualization. Aggregate RGB SNR is the mean of the recording-level red-, green-, and blue-channel SNR measures used in the sensitivity analysis. The near-zero correlation indicates that this measured aggregate SNR metric has little linear association with $D_{HR}$ in these data; it is not interpreted as a causal test of optical mechanisms or as evidence that optical factors are unimportant.}}
\label{fig:snrhr}
\end{figure}

Measured channel SNR was low across the dataset and varied only modestly
by Fitzpatrick group. As summarized in Table~\ref{tab:snr}, group-mean values for the red, green, and blue
channels remained near -15 to -16 dB, with approximately 0.67 dB of
aggregate variation across Fitzpatrick groups. The channel-specific distributions underlying these summaries are shown in Supplementary Figure~\ref{fig:s2}. This small SNR difference
contrasted with the 9.21 beats/min increase in mean DHR from Fitzpatrick
III to VI. Across the 625-record Fitzpatrick analysis set, recording-level aggregate RGB SNR was essentially uncorrelated with absolute HR discrepancy (Pearson r=0.014, p=0.718; Figure~\ref{fig:snrhr}).

\begin{longtable}{@{}
 >{\raggedright\arraybackslash}p{(\columnwidth - 8\tabcolsep) * \real{0.2000}}
 >{\raggedright\arraybackslash}p{(\columnwidth - 8\tabcolsep) * \real{0.2000}}
 >{\raggedright\arraybackslash}p{(\columnwidth - 8\tabcolsep) * \real{0.2000}}
 >{\raggedright\arraybackslash}p{(\columnwidth - 8\tabcolsep) * \real{0.2000}}
 >{\raggedright\arraybackslash}p{(\columnwidth - 8\tabcolsep) * \real{0.2000}}@{}}

\multicolumn{5}{@{}p{\linewidth}@{}}{\textbf{Table~\thetable{} \textbar{} Measured recording-level RGB SNR by Fitzpatrick group (dB).}\par\vspace{2pt}\textit{Recording-level red-, green-, and blue-channel SNR values and their aggregate mean are summarized by Fitzpatrick group for the 625-record analysis set. These values are descriptive signal-quality measures and are not interpreted as direct estimates of melanin concentration or tissue optical absorption.}} \\[4pt]
\toprule\noalign{}
\begin{minipage}[b]{\linewidth}\raggedright
\textbf{Fitzpatrick}
\end{minipage} & \begin{minipage}[b]{\linewidth}\raggedright
\textbf{N}
\end{minipage} & \begin{minipage}[b]{\linewidth}\raggedright
\textbf{SNR\_R}
\end{minipage} & \begin{minipage}[b]{\linewidth}\raggedright
\textbf{SNR\_G}
\end{minipage} & \begin{minipage}[b]{\linewidth}\raggedright
\textbf{SNR\_B}
\end{minipage} \\
\midrule\noalign{}
\endhead
\bottomrule\noalign{}
\endlastfoot
\label{tab:snr}III & 338 & -15.55 & -15.51 & -15.22 \\
IV & 77 & -15.96 & -15.85 & -15.64 \\
V & 120 & -16.52 & -16.02 & -15.76 \\
VI & 120 & -15.97 & -16.02 & -16.26 \\
\end{longtable}

\subsubsection{A simple Beer--Lambert-inspired attenuation adequacy check does not reproduce
measured
SNR}\label{a-simple-beer-lambert-attenuation-model-fails-to-reproduce-measured-snr}

\begin{figure}[!htbp]
\centering
\includegraphics[width=\textwidth]{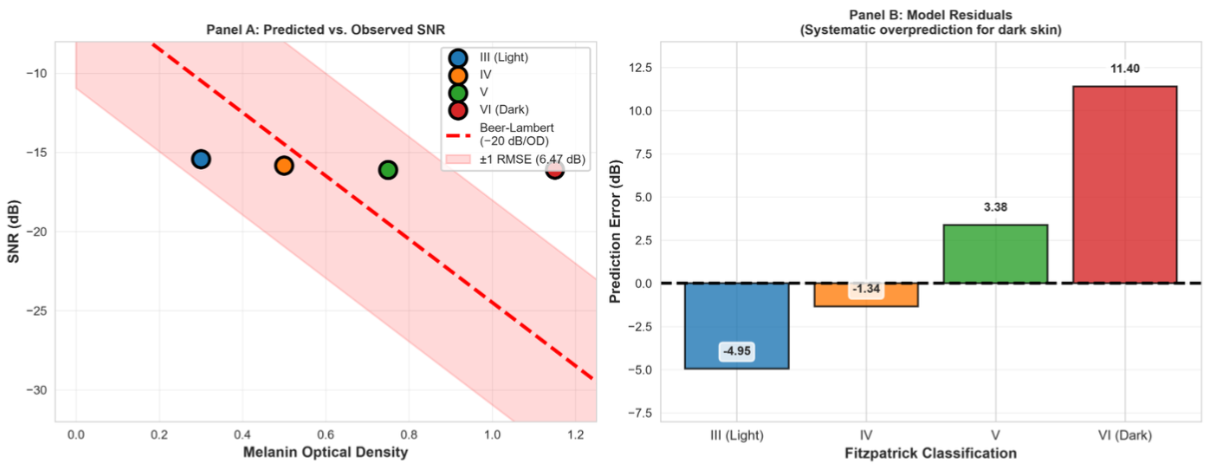}
\caption{\textbf{Predicted versus observed SNR and model residuals across Fitzpatrick groups.} \textit{(A) Observed aggregate RGB SNR is compared with the monotonic pattern predicted by the deliberately simplified Beer--Lambert-inspired relation $\mathrm{SNR}_{dB}=\mathrm{baseline\ SNR}-20\,OD_{melanin}$. The model RMSE is 6.47 dB. (B) Group-level prediction residuals show increasing positive error toward the darker-skin groups, with the largest discrepancy for Fitzpatrick VI. Residuals are observed minus predicted SNR, so increasingly positive residuals indicate that the simple model predicts substantially more attenuation than was measured. This adequacy check shows that the specified one-parameter monotonic attenuation pattern does not reproduce the group-level SNR observations. The model is not a physiological melanin estimator, does not represent the full camera/illumination/tissue imaging chain, and is not used to correct rPPG signals or establish causality.}}
\label{fig:beerlambert}
\end{figure}

We next evaluated a deliberately simplified Beer-Lambert-inspired
attenuation model, motivated by the wavelength-dependent absorption and scattering of biological tissue and melanin \cite{jacques2013optical}, SNRdB = baseline SNR - 20 \ensuremath{\times} ODmelanin, as an adequacy
check rather than as a complete optical model. Table~\ref{tab:beer} reports the observed-versus-predicted comparison. The predicted SNR
decreased sharply from Fitzpatrick III to VI, whereas the measured
pattern was nearly flat; the resulting RMSE was 6.47 dB. The tested
attenuation model therefore did not reproduce the observed SNR pattern
and was not used for correction or causal inference. Figure~\ref{fig:beerlambert}A shows the predicted-versus-observed mismatch, while Figure~\ref{fig:beerlambert}B shows the group-level residual pattern.

\begin{longtable}{@{}
 >{\raggedright\arraybackslash}p{(\columnwidth - 6\tabcolsep) * \real{0.2500}}
 >{\raggedright\arraybackslash}p{(\columnwidth - 6\tabcolsep) * \real{0.2500}}
 >{\raggedright\arraybackslash}p{(\columnwidth - 6\tabcolsep) * \real{0.2500}}
 >{\raggedright\arraybackslash}p{(\columnwidth - 6\tabcolsep) * \real{0.2500}}@{}}

\multicolumn{4}{@{}p{\linewidth}@{}}{\textbf{Table~\thetable{} \textbar{} Observed versus predicted SNR for the simplified Beer--Lambert-inspired adequacy check.}\par\vspace{2pt}\textit{Observed group-level aggregate RGB SNR is compared with the monotonic attenuation pattern generated by the simplified Beer--Lambert-inspired relation. Residuals quantify observed-minus-predicted SNR and are used only to assess adequacy of the specified pattern, not to estimate melanin or establish a causal optical mechanism.}} \\[4pt]
\toprule\noalign{}
\begin{minipage}[b]{\linewidth}\raggedright
\textbf{Fitzpatrick}
\end{minipage} & \begin{minipage}[b]{\linewidth}\raggedright
\textbf{Observed SNR}
\end{minipage} & \begin{minipage}[b]{\linewidth}\raggedright
\textbf{Predicted SNR}
\end{minipage} & \begin{minipage}[b]{\linewidth}\raggedright
\textbf{Prediction error}
\end{minipage} \\
\midrule\noalign{}
\endhead
\bottomrule\noalign{}
\endlastfoot
\label{tab:beer}III & -15.43 & -10.48 & -4.94 \\
IV & -15.82 & -14.48 & -1.33 \\
V & -16.10 & -19.48 & +3.39 \\
VI & -16.08 & -27.48 & +11.40 \\
\end{longtable}

This negative result does not identify the source of the endpoint
disparity. Optical acquisition, camera response and compression, ROI
selection, color-channel interactions, rPPG extraction, and
interactions among these factors remain plausible contributors.

\subsubsection{Measured-SNR adjustment leaves the Fitzpatrick coefficient
unchanged}\label{measured-snr-adjustment-leaves-the-fitzpatrick-coefficient-unchanged}

The SNR-adjusted sensitivity model (Model 1b) added a single
aggregate covariate, $\mathrm{SNR}_{\mathrm{RGB}}=(\mathrm{SNR}_{R}+\mathrm{SNR}_{G}+\mathrm{SNR}_{B})/3$, to Model 1. The final coefficients are reported in Table~\ref{tab:model1b}. $\mathrm{SNR}_{\mathrm{RGB}}$ was
not associated with $D_{HR}$ (0.0769 beats/min/dB, 95\% CI -0.2283 to 0.3821;
p=0.621). More importantly, the Fitzpatrick VI-versus-III coefficient
changed only from +9.3218 to +9.3678 beats/min, corresponding to -0.49\%
descriptive coefficient attenuation. Adjustment for measured aggregate
RGB SNR therefore produced essentially no attenuation of the
Fitzpatrick-associated endpoint discrepancy. The near-invariance of the Fitzpatrick coefficients before and after SNR adjustment is also shown in Supplementary Figure~\ref{fig:s3}.

{\footnotesize
\setlength{\tabcolsep}{1.5pt}
\begin{longtable}{@{}
 >{\raggedright\arraybackslash}p{(\columnwidth - 8\tabcolsep) * \real{0.2000}}
 >{\raggedright\arraybackslash}p{(\columnwidth - 8\tabcolsep) * \real{0.2000}}
 >{\raggedright\arraybackslash}p{(\columnwidth - 8\tabcolsep) * \real{0.2000}}
 >{\raggedright\arraybackslash}p{(\columnwidth - 8\tabcolsep) * \real{0.2000}}
 >{\raggedright\arraybackslash}p{(\columnwidth - 8\tabcolsep) * \real{0.2000}}@{}}

\multicolumn{5}{@{}p{\linewidth}@{}}{\textbf{Table~\thetable{} \textbar{} SNR-adjusted sensitivity model.}\par\vspace{2pt}\textit{Model 1b adds recording-level aggregate RGB SNR to the subject-aware endpoint-discrepancy model while retaining Fitzpatrick group, lighting, motion/activity, and the subject random intercept. The table evaluates whether SNR adjustment materially changes the Fitzpatrick-associated coefficients.}} \\[4pt]
\toprule\noalign{}
\begin{minipage}[b]{\linewidth}\raggedright
\textbf{Contrast}
\end{minipage} & \begin{minipage}[b]{\linewidth}\raggedright
\textbf{Model 1}
\end{minipage} & \begin{minipage}[b]{\linewidth}\raggedright
\textbf{Model 1b + SNR\_RGB}
\end{minipage} & \begin{minipage}[b]{\linewidth}\raggedright
\textbf{Model 1b 95\% CI}
\end{minipage} & \begin{minipage}[b]{\linewidth}\raggedright
\textbf{Model 1b p}
\end{minipage} \\
\midrule\noalign{}
\endhead
\bottomrule\noalign{}
\endlastfoot
\label{tab:model1b}IV vs III & 6.37 & 6.40 & {[}-0.08, 12.89{]} & 0.053 \\
V vs III & 6.72 & 6.77 & {[}1.25, 12.30{]} & 0.016 \\
VI vs III & 9.32 & 9.37 & {[}3.83, 14.90{]} & 0.001 \\
SNR\_RGB &, & 0.077 bpm/dB & {[}-0.228, 0.382{]} & 0.621 \\
\end{longtable}
}

\begin{figure}[!htbp]
\centering
\includegraphics[width=0.88\textwidth]{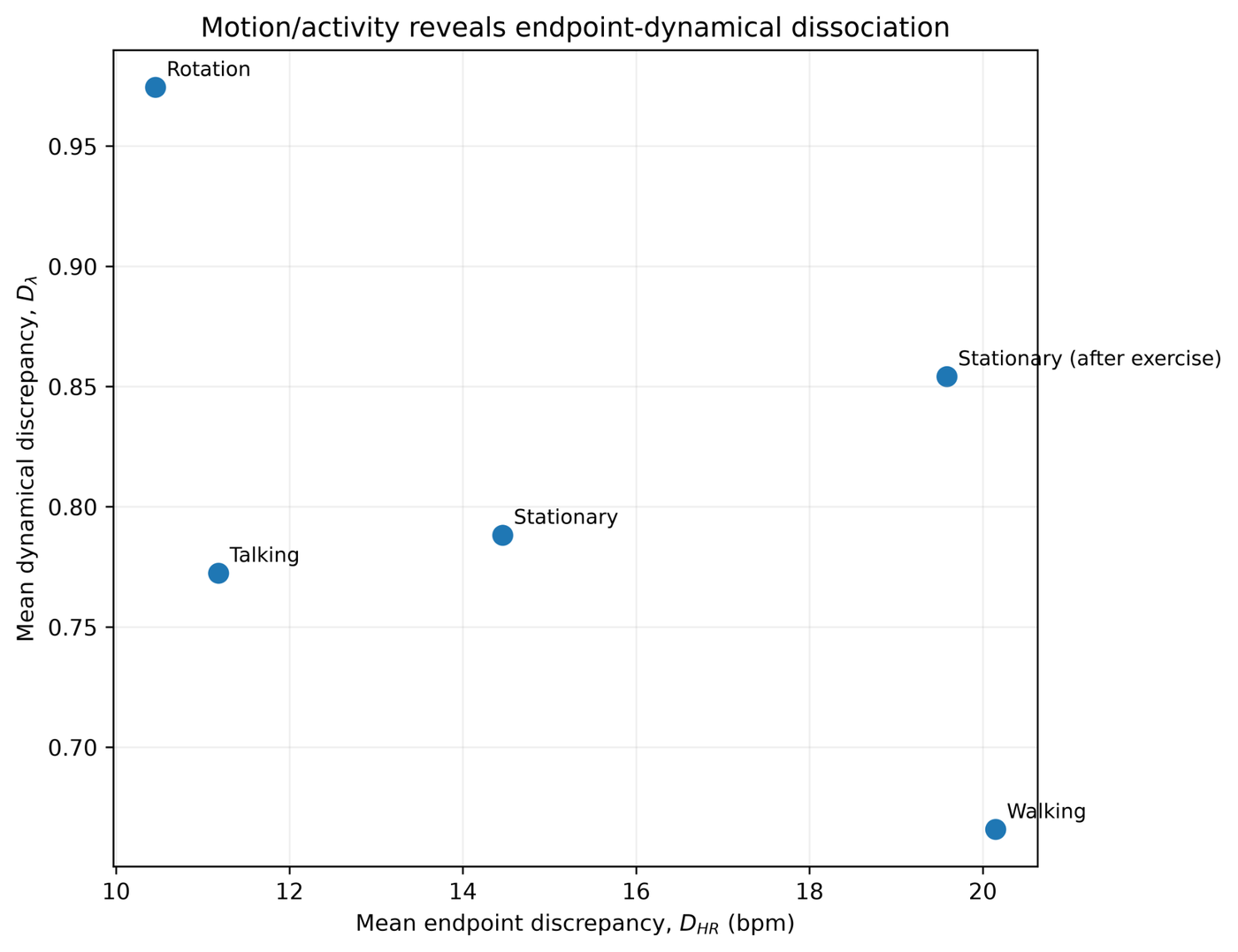}
\caption{\textbf{Motion/activity reveals endpoint--dynamical dissociation.} \textit{Each point represents the condition-specific mean endpoint discrepancy $D_{HR}$ and mean dynamical discrepancy $D_{\lambda}$. Conditions with similar or larger endpoint discrepancy do not necessarily exhibit correspondingly larger $D_{\lambda}$; for example, walking has high mean endpoint discrepancy but the lowest mean dynamical discrepancy among the displayed conditions. Because $D_{HR}$ and $D_{\lambda}$ have different units and scales, the figure evaluates condition ordering and co-variation rather than equality of numerical magnitude. The plot is descriptive and illustrates that acquisition conditions can affect endpoint and dynamical recoverability differently.}}
\label{fig:motiondissociation}
\end{figure}

The condition-level endpoint--dynamical dissociation is summarized graphically in Figure~\ref{fig:motiondissociation}; notably, walking combines high mean endpoint discrepancy with the lowest mean dynamical discrepancy among the displayed motion/activity conditions.

\subsection{Observation context partially reduces held-out endpoint discrepancy}\label{condition-aware-gap-results}

The subject-held-out experiment tested the extent to which measured observation state reduced the PPG--rPPG endpoint discrepancy, with MAE as the primary performance measure. On the 625-record condition-aware analysis subset, raw camera-derived rPPG HR had MAE 15.24 bpm, compared with 15.26 bpm in the full 655-record fixed-pathway benchmark. A linear learner using rPPG HR alone ($M_0$) reduced MAE to 12.54 bpm. Thus, $M_0$ is a learned calibration baseline; comparisons with $M_1$--$M_4$ quantify the incremental MAE reduction attributable to observation context beyond that calibration rather than comparing contextual models directly with the raw 15.24-bpm error. Adding motion ($M_1$) reduced MAE to 11.15 bpm, and adding lighting ($M_2$) reduced MAE further to 10.87 bpm.

The same incremental pattern was observed across all three learner families. $M_2$ produced the lowest MAE for linear regression (10.87 bpm), ridge regression (10.87 bpm), and random forest (11.23 bpm), corresponding to reductions of 13.32\%, 13.32\%, and 12.77\% relative to their respective $M_0$ baselines. The subject-bootstrap 95\% CIs for the $M_2-M_0$ MAE difference were entirely below zero for linear regression ($-2.36$ to $-0.98$ bpm), ridge regression ($-2.35$ to $-0.99$ bpm), and random forest ($-2.38$ to $-0.93$ bpm). Adding Fitzpatrick group ($M_3$) or aggregate RGB SNR ($M_4$) did not further reduce the observed MAE in any learner family. Thus, the predictive contribution of motion and lighting was consistent across linear, regularized-linear, and nonlinear learners rather than being specific to a single model class.

Table~\ref{tab:condition_aware_learners} summarizes the subject-held-out MAE across the five nested information sets and all three learner families.

\begin{table}[!htbp]
\centering
\caption{\textbf{Subject-held-out endpoint prediction across learner families.} \textit{Out-of-fold MAE (bpm) is reported for each nested information set and learner family. Percentages in parentheses denote MAE reduction relative to the corresponding $M_0$ baseline within the same learner family. $M_0$: rPPG HR only; $M_1$: +motion; $M_2$: +lighting; $M_3$: +Fitzpatrick group; $M_4$: +aggregate RGB SNR.}}
\label{tab:condition_aware_learners}
\small
\resizebox{0.88\linewidth}{!}{
\begin{tabular}{lccc}
\toprule
\textbf{Information set} & \textbf{Linear} & \textbf{Ridge} & \textbf{Random forest} \\
\midrule
$M_0$ & 12.54 (0.00\%) & 12.54 (0.00\%) & 12.88 (0.00\%) \\
$M_1$ & 11.15 (11.14\%) & 11.15 (11.13\%) & 11.48 (10.87\%) \\
$M_2$ & \textbf{10.87 (13.32\%)} & \textbf{10.87 (13.32\%)} & \textbf{11.23 (12.77\%)} \\
$M_3$ & 11.16 (11.02\%) & 11.15 (11.15\%) & 11.65 (9.53\%) \\
$M_4$ & 11.23 (10.45\%) & 11.22 (10.58\%) & 11.89 (7.69\%) \\
\bottomrule
\end{tabular}
}
\end{table}

Adding Fitzpatrick group did not further improve held-out performance: linear-model MAE increased from 10.87 bpm in $M_2$ to 11.16 bpm in $M_3$, and the bootstrap CI for improvement relative to $M_0$ crossed zero at its upper bound. Adding aggregate RGB SNR increased MAE further to 11.23 bpm in $M_4$, again with a bootstrap interval crossing zero. The same ordering appeared in ridge and random-forest models. These results distinguish factors associated with heterogeneity from factors that improve prediction: motion and lighting provided useful observation context, whereas Fitzpatrick category and aggregate RGB SNR did not add generalizable predictive benefit after motion and lighting were already included.

\begin{figure}[!htbp]
\centering
\includegraphics[width=\textwidth]{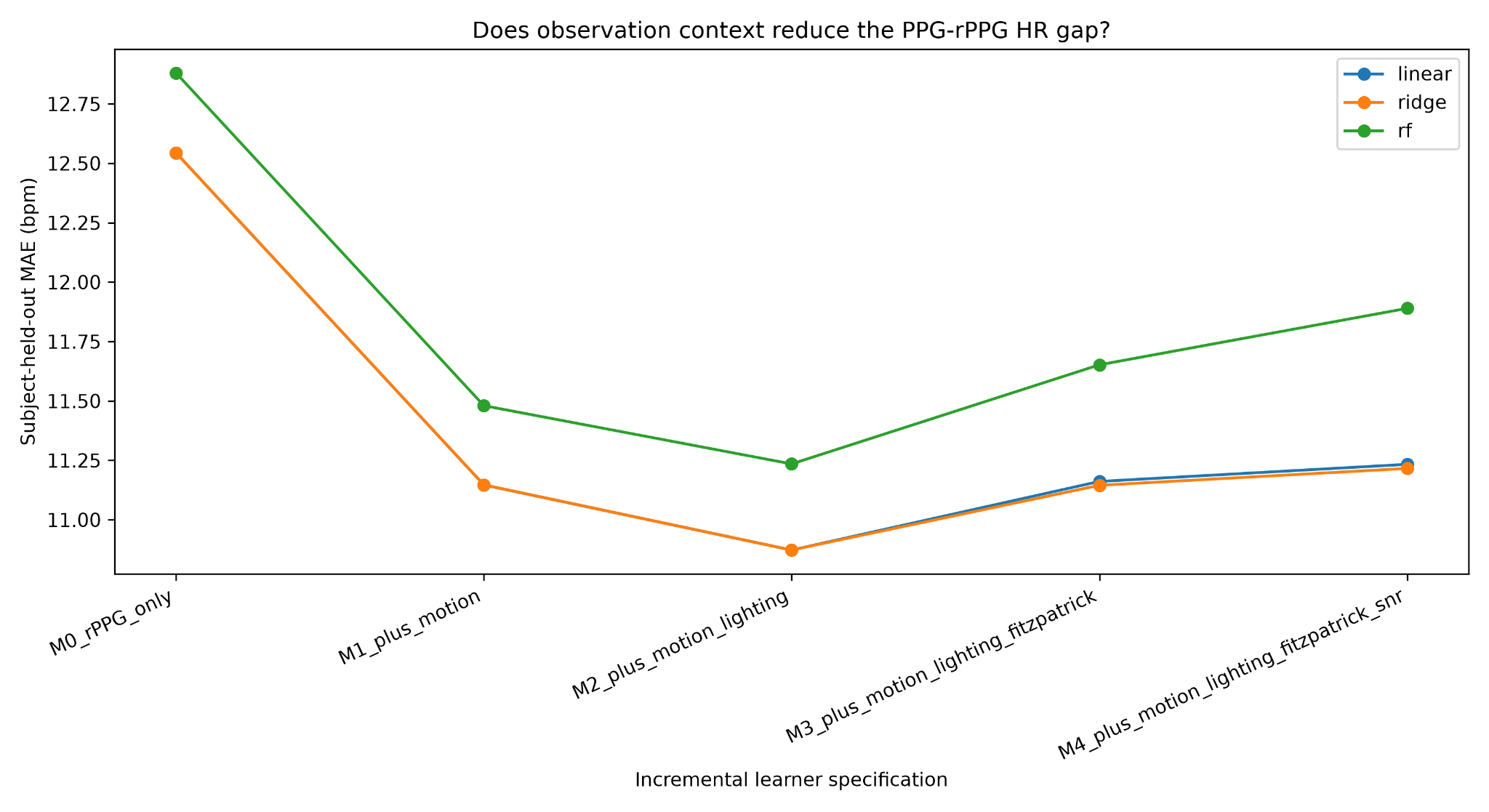}
\caption{\textbf{Subject-held-out reduction of the PPG--rPPG HR gap using measured observation context.} \textit{Out-of-fold MAE is shown for linear, ridge, and random-forest learners as predictors are added incrementally: rPPG HR only ($M_0$), +motion ($M_1$), +lighting ($M_2$), +Fitzpatrick group ($M_3$), and +aggregate RGB SNR ($M_4$). Motion and lighting consistently reduce MAE, with the best result at $M_2$; adding Fitzpatrick group or aggregate SNR does not improve the best model. The analysis uses the 625-record, 32-subject condition-aware subset; its raw camera-derived rPPG HR MAE is 15.24 bpm, compared with 15.26 bpm in the full 655-record benchmark. Five-fold GroupKFold validation is subject-held-out, so recordings from a participant do not appear in both training and test folds. For the linear learner, MAE decreases from 12.54 bpm at $M_0$ to 10.87 bpm at $M_2$ (13.32\% reduction; subject-bootstrap 95\% CI for the MAE difference, -2.36 to -0.98 bpm). The remaining error is substantial and does not establish equivalence between rPPG and contact PPG.}}
\label{fig:gapreduction}
\end{figure}

The corresponding predicted-versus-reference plot (Supplementary Figure~\ref{fig:s5}) shows substantial regression toward the central HR range despite the MAE improvement. Accordingly, this experiment supports only partial narrowing of the endpoint gap. It does not show that camera-derived rPPG can be used interchangeably with contact PPG or that observation context reconstructs the source waveform.

\section{Discussion}\label{discussion}

\subsection{Property-specific preservation across an observation pathway}

The central signal-processing finding is that PPG-to-rPPG conversion does not have a single level of recoverability. Under the same fixed CHROM-based observation pathway, endpoint HR, conventional temporal and spectral structure, and maximal-Lyapunov dynamics exhibited different degrees of recording-specific correspondence. Temporal autocorrelation retained a modest matched-pair advantage, whereas spectral and recurrence-rate measures showed little separation from shuffled controls, and recording-specific $\lambda_{\max}$ correspondence was not detectable. Recoverability should therefore be assigned to a defined physiological property under a defined observation pathway rather than globally to the rPPG modality. Equivalently, the relevant object is the recoverability of a defined property $\phi_j$ under an observation transformation $\mathcal{H}_{\theta}$, rather than recoverability of the derived modality as a whole.

These findings refine our earlier work on skin-tone-robust topological signal processing \cite{oladunni2026skin}. Robustness of a topological descriptor across population or observation groups does not by itself establish that the corresponding property is preserved recording by recording across heterogeneous sensing pathways.

The nonlinear analysis provides the clearest counterexample to inference from population-level plausibility. PPG and rPPG $\lambda_{\max}$ estimates occupied overlapping ranges, yet matched-recording correspondence was essentially indistinguishable from the recording-identity null and rPPG variability was compressed. Thus, a descriptor can appear physiologically plausible at the population level without preserving which recording had which value. This result is consistent with state-space and Lyapunov methods, for which the estimated dynamical quantity is tied to the reconstructed trajectory rather than to marginal population overlap alone \cite{takens1981detecting,rosenstein1993practical}. In this manuscript, failure of matched correspondence is interpreted as evidence against preservation of the tested property; it is not a formal information-theoretic estimate of information loss.

The conventional structural controls reinforce the same distinction. ACF similarity retained modest recording specificity, whereas PSD and recurrence-rate differences provided little matched-versus-shuffled separation and determinism was non-discriminative under the implemented parameterization. The correspondence-breaking control is therefore important for distinguishing recording-specific preservation from structure that can also arise in nonmatched signals. This use of matched identity is an analysis design choice of the present study rather than an established rPPG benchmark convention.

\subsection{Observation-associated heterogeneity is also property-specific}

Observation conditions did not affect all recovered properties in parallel. Endpoint discrepancy varied substantially with motion/activity and Fitzpatrick group after simultaneous subject-aware adjustment, whereas the tested dynamical discrepancy showed a different pattern. The Fitzpatrick-by-motion block provided secondary evidence that endpoint heterogeneity may depend jointly on subject-associated and acquisition conditions, but the smaller Fitzpatrick strata contain few independent participants; this interaction therefore motivates replication rather than a definitive characterization of subgroup-specific motion effects. The corresponding Fitzpatrick-by-lighting interaction was not supported, and neither interaction was supported for $D_{\lambda}$.

The Fitzpatrick-associated endpoint contrast persisted after adjustment for the measured acquisition covariates, and adding aggregate RGB SNR did not materially account for that heterogeneity. This does not establish that optical signal quality is irrelevant. Rather, the simple pathway
\[
\text{Fitzpatrick group}\rightarrow\text{aggregate RGB SNR}\rightarrow D_{HR}
\]
was not supported by the specified analyses. Aggregate RGB SNR may not capture wavelength-dependent tissue optics, camera response, spatial ROI effects, channel mixing, compression, or algorithm-by-acquisition interactions that contribute to rPPG formation \cite{jacques2013optical,setchfield2024skinoptics,wang2017algorithmic}.

The optical-attenuation adequacy check narrows the mechanism further without identifying it. The simple Beer--Lambert-inspired model predicted a substantially stronger Fitzpatrick-dependent SNR decline than was measured. The appropriate conclusion is therefore that this simplified monotonic attenuation model is inadequate for reproducing the observed SNR pattern, not that Beer--Lambert physics or tissue optics are unimportant. The end-to-end camera pathway contains coupled optical, camera, behavioral, spatial-sampling, and algorithmic transformations that were not separately manipulated here.

The pattern for $D_{\lambda}$ further illustrates why heterogeneity should not be generalized from one property to another. Fitzpatrick- and lighting-associated differences were not detected for this discrepancy, while several motion/activity contrasts differed from the reference condition. Smaller $D_{\lambda}$ values under particular conditions cannot be interpreted as improved dynamical preservation because recording-specific $\lambda_{\max}$ correspondence was absent overall. The relevant result is the dissociation: factors associated with endpoint discrepancy do not map monotonically onto the tested dynamical discrepancy. A condition associated with degradation of one recovered property should therefore not automatically be assumed to degrade another property in the same way.

\subsection{Diagnostic association does not imply predictive actionability}

The condition-aware experiment asks a different question from the mixed-effects analysis: whether measured observation context can improve prediction of contact-PPG HR for unseen subjects. The rPPG-HR-only model ($M_0$) serves as a learned calibration baseline, so subsequent improvements quantify the incremental value of context beyond systematic scale and offset correction. Motion and lighting reduced held-out error beyond this baseline, whereas adding Fitzpatrick group or aggregate RGB SNR did not improve the best model. Ridge and random forest reproduced the same ordering and did not outperform the linear specification.

This separation between inferential association and predictive utility is important. A variable can identify where discrepancy differs without providing incremental predictive information once other observation context is available. Conversely, motion and lighting were useful for partial endpoint-gap reduction even though the remaining error and compressed prediction range were substantial. The experiment therefore supports condition-aware gap reduction, not reconstruction of the source PPG waveform, physiological correction, or interchangeability of camera-derived rPPG with contact PPG.

\subsection{Scope and relation to prior representation frameworks}

The weak CHROM endpoint benchmark does not establish an intrinsic limit of camera-based physiological sensing. CHROM was used as an established, unsupervised observation pathway whose benchmark behavior on MMPD could be reproduced without selecting an rPPG method according to downstream results. The conclusions are therefore operator-specific: they characterize which tested properties remain accessible under this CHROM-based transformation, not what is fundamentally unrecoverable by every possible camera-rPPG method.

The present findings also provide an observation-level extension of the representation question introduced in the prior CFD and CST work discussed in Section~\ref{introduction}. CFD motivates the broader premise that physiological information available for inference can depend on representation, while CST emphasizes temporal and dynamical organization in contact PPG. The present study asks the next observation-level question: when the sensing pathway changes from contact PPG to camera-derived rPPG, which tested properties retain recording-specific correspondence? This relationship is conceptual rather than assumptive; the current framework and empirical conclusions stand independently of CFD or CST \cite{oladunni2025cfd,oladunni2026cst}.

Taken together, the results show that endpoint discrepancy, conventional structural correspondence, maximal-Lyapunov correspondence, measured RGB SNR, and acquisition-condition effects do not collapse to a single recoverability axis. Camera-derived physiological validation should therefore specify the property claimed to be recovered, test recording-specific correspondence for that property when appropriate, and evaluate its stability across relevant observation conditions.

\section{Limitations}\label{limitations}

Several limitations define the scope of these findings. First, all analyses used MMPD. Although the dataset contains hundreds of recordings spanning multiple acquisition conditions, the number of independent participants is much smaller, particularly within some Fitzpatrick strata. Subject-aware mixed-effects models, grouped cross-validation, and subject-level bootstrap resampling address repeated measurements but do not substitute for independent external validation. The Fitzpatrick-by-motion finding should therefore be treated as replication-worthy evidence rather than a definitive characterization of subgroup-specific motion effects.

Second, the study intentionally used one fixed, established CHROM-based camera-rPPG observation pathway. The observed endpoint, structural, and dynamical correspondence therefore characterize this operator on MMPD and do not establish a fundamental limit of camera-based physiological sensing. Alternative rPPG extraction methods, acquisition systems, or learned observation models may preserve different properties and should be evaluated using the same recording-specific framework.

Third, the tested properties--HR, autocorrelation, spectral structure, recurrence measures, and maximal Lyapunov exponent--are representative rather than exhaustive. The observed behavior of these properties does not determine the behavior of every physiological property that could be extracted from PPG or rPPG. Moreover, each property is observed through a finite-sample estimator. In particular, $\lambda_{\max}$ depends on preprocessing, state-space reconstruction, validity criteria, and estimator behavior \cite{takens1981detecting,rosenstein1993practical}. The null $\lambda_{\max}$ correspondence result therefore applies to the implemented dynamical representation and should not be generalized to all nonlinear cardiovascular structure.

Fourth, the observation-condition analyses are associational rather than causal. Fitzpatrick category is not a direct measure of melanin concentration or tissue optical properties \cite{weir2024skintone}, and aggregate RGB SNR is not a complete description of camera-domain information quality \cite{elgendi2024sqi}. Likewise, the Beer--Lambert-inspired analysis was a deliberately simplified adequacy check rather than a full physical model of camera-rPPG formation. The persistence of Fitzpatrick-associated endpoint heterogeneity after RGB-SNR adjustment therefore neither identifies a causal mechanism nor establishes that optical signal quality is irrelevant. Direct optical measurements, controlled acquisition, and prospective designs that separate subject-associated from acquisition factors are needed for mechanistic attribution.

Finally, the condition-aware experiment was designed primarily to test whether measured observation context reduces subject-held-out MAE in the PPG--rPPG endpoint discrepancy, rather than to maximize explained variance. Motion and lighting produced consistent MAE reductions across all three learner families, with $M_2$ reducing MAE by 13.32\%, 13.32\%, and 12.77\% relative to $M_0$ for linear regression, ridge regression, and random forest, respectively. However, the maximum out-of-fold $R^2$ was 0.189, indicating that the measured observation descriptors account for only a limited portion of recording-level variation. Thus, the RQ5 findings support partial reduction of endpoint error as measured by MAE, rather than complete explanation or correction of the PPG--rPPG discrepancy. The experiment also predicts contact-PPG HR rather than reconstructing the contact-PPG waveform and therefore does not establish physiological waveform recovery or interchangeability of camera-derived rPPG with contact PPG. Future validation should extend the framework across independent datasets, multiple rPPG operators and sensing systems, direct optical measurements, and additional physiological property operators.

\section{Conclusion}\label{conclusion}

This study supports a property-specific framework for evaluating physiological measurements across heterogeneous observation pathways. Under the evaluated contact-PPG-to-camera-rPPG transformation, population-level plausibility did not guarantee recording-specific preservation, and observation-associated degradation was neither uniform across properties nor completely captured by simple signal-quality descriptors. These findings characterize the evaluated observation pathway rather than a fundamental limit of camera sensing.

More broadly, cross-modal physiological validation should identify the property claimed to be recovered and establish its recording-level preservation and stability under relevant observation conditions rather than infer signal equivalence from endpoint performance alone.

\section{Data availability}\label{data-availability}

The source MMPD dataset is available from the dataset authors under its
stated access conditions. Derived matched-analysis tables and
exclusion/accounting records will be made available with the accepted
article subject to the original dataset terms.

\section{Code availability}\label{code-availability}

The analysis code used in this study will be made publicly available in a public repository upon publication of the article.

\section*{Supplementary figures}\label{supplementary-figures}
\setcounter{figure}{0}
\renewcommand{\thefigure}{S\arabic{figure}}
\renewcommand{\theHfigure}{S\arabic{figure}}

\begin{figure}[!htbp]
\centering
\includegraphics[width=0.9\textwidth]{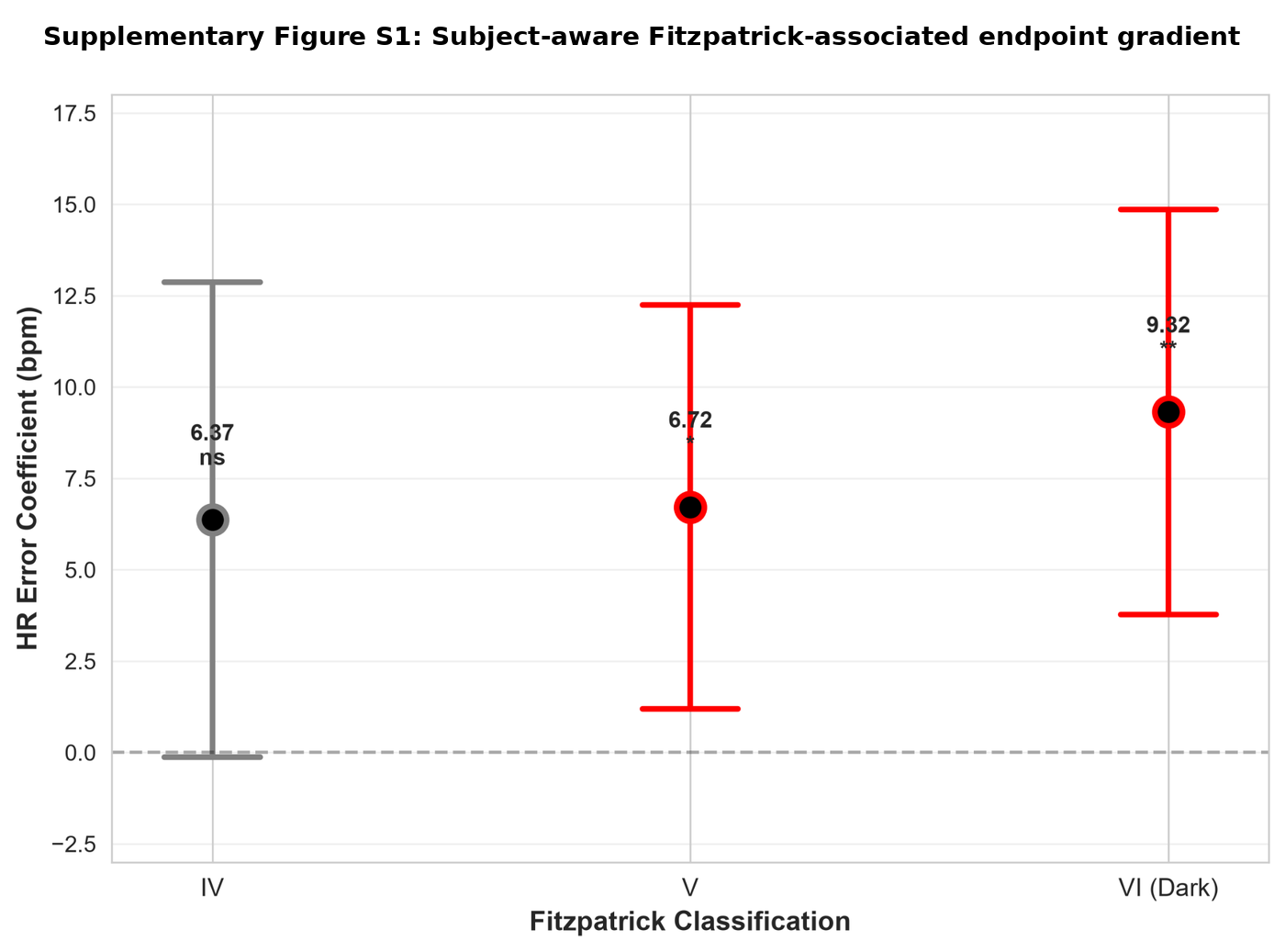}
\caption{\textbf{Subject-aware Fitzpatrick-associated endpoint gradient.} \textit{Mixed-effects coefficients for Fitzpatrick IV, V, and VI relative to Fitzpatrick III after adjustment for lighting and motion/activity with a subject random intercept ($N=625$ recordings, 32 subjects). Coefficients are expressed in beats/min and use Fitzpatrick III as the reference group; error bars denote 95\% confidence intervals. This coefficient-focused display complements the combined descriptive-and-adjusted presentation in Figure~\ref{fig:heterogeneous}.}}
\label{fig:s1}
\end{figure}

\begin{figure}[!htbp]
\centering
\includegraphics[width=\textwidth]{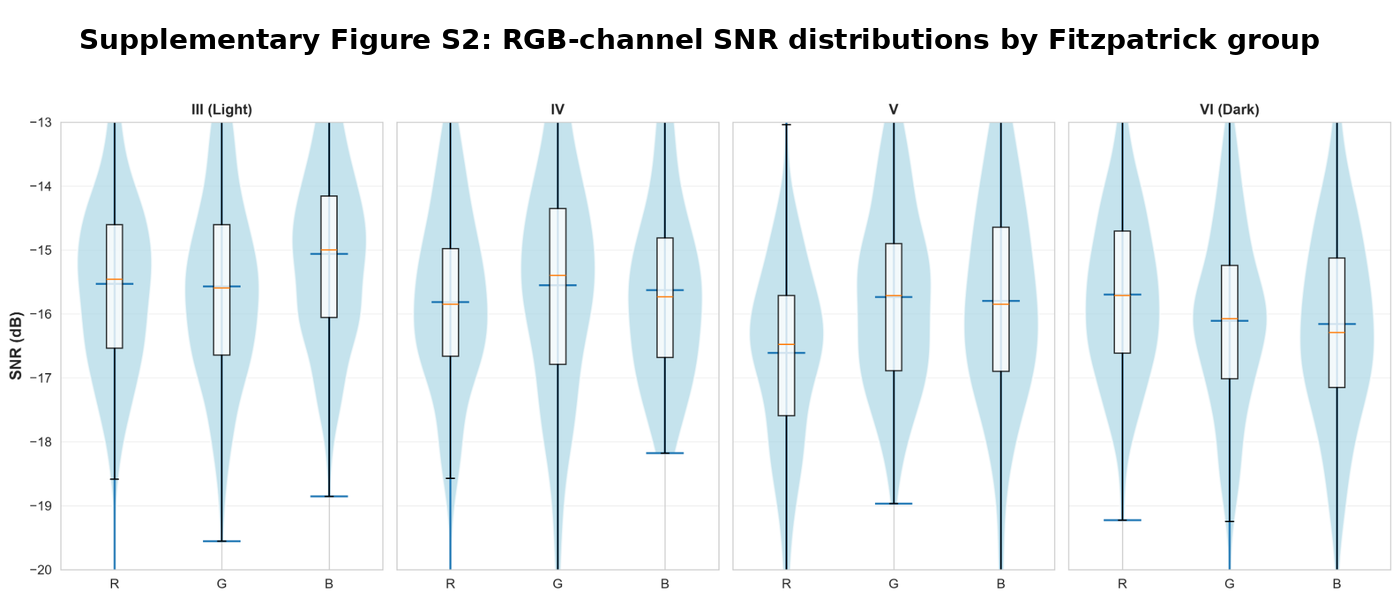}
\caption{\textbf{RGB-channel SNR distributions by Fitzpatrick group.} \textit{Recording-level red-, green-, and blue-channel SNR distributions are shown for the 625-record Fitzpatrick analysis set. The extensive overlap and comparatively small shifts in central tendency provide the distributional context for Table~\ref{tab:snr}; these measurements are descriptive signal-quality summaries and are not direct measures of melanin concentration or tissue optical absorption.}}
\label{fig:s2}
\end{figure}

\begin{figure}[!htbp]
\centering
\includegraphics[width=0.9\textwidth]{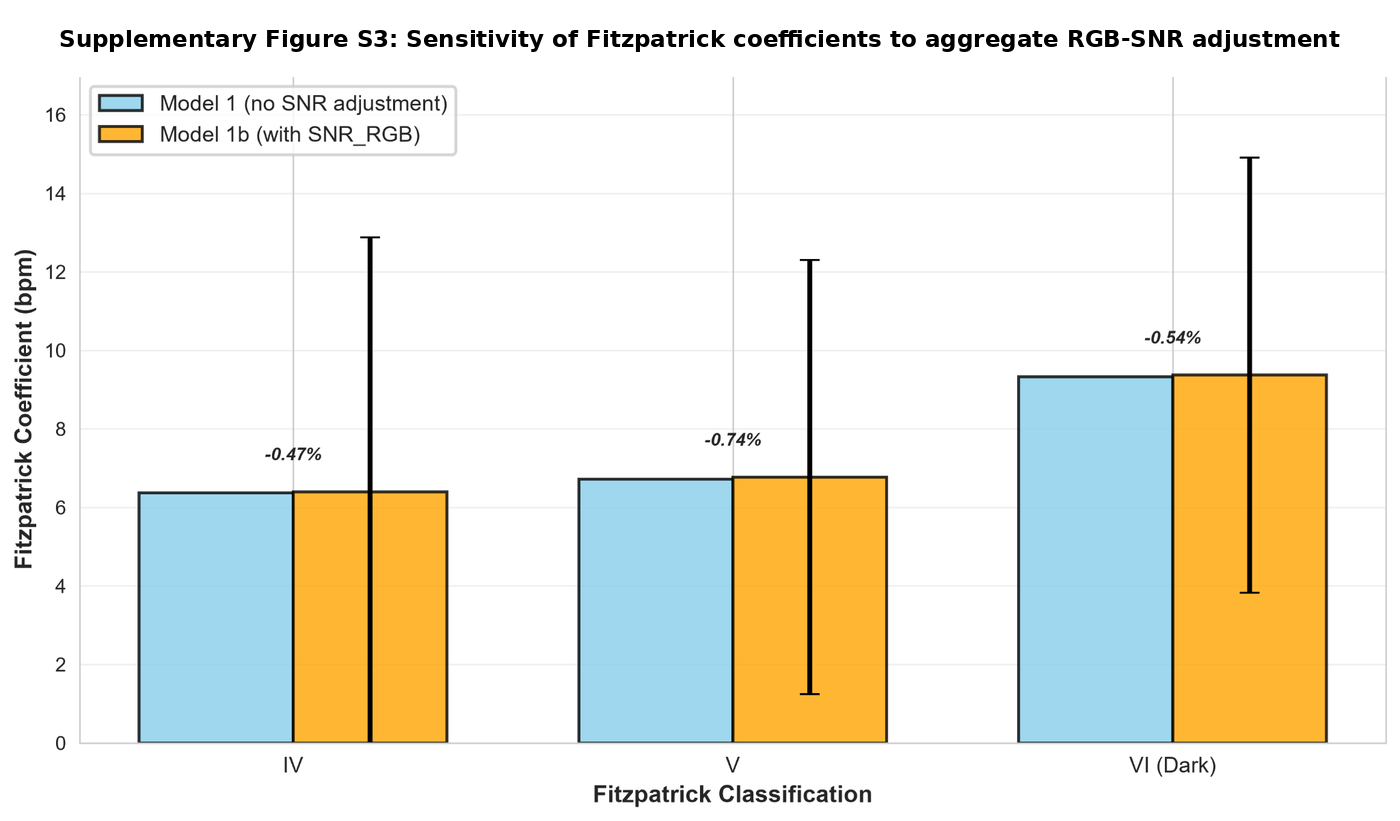}
\caption{\textbf{Sensitivity of Fitzpatrick coefficients to aggregate RGB-SNR adjustment.} \textit{Model 1b adds measured aggregate RGB SNR to Model 1 while retaining Fitzpatrick group, lighting, motion/activity, and the subject random intercept. The displayed Fitzpatrick contrasts change by less than 1\% relative to Model 1, indicating negligible descriptive attenuation after SNR adjustment. This is a sensitivity analysis and is not interpreted as causal mediation.}}
\label{fig:s3}
\end{figure}

\begin{figure}[!htbp]
\centering
\includegraphics[width=\textwidth]{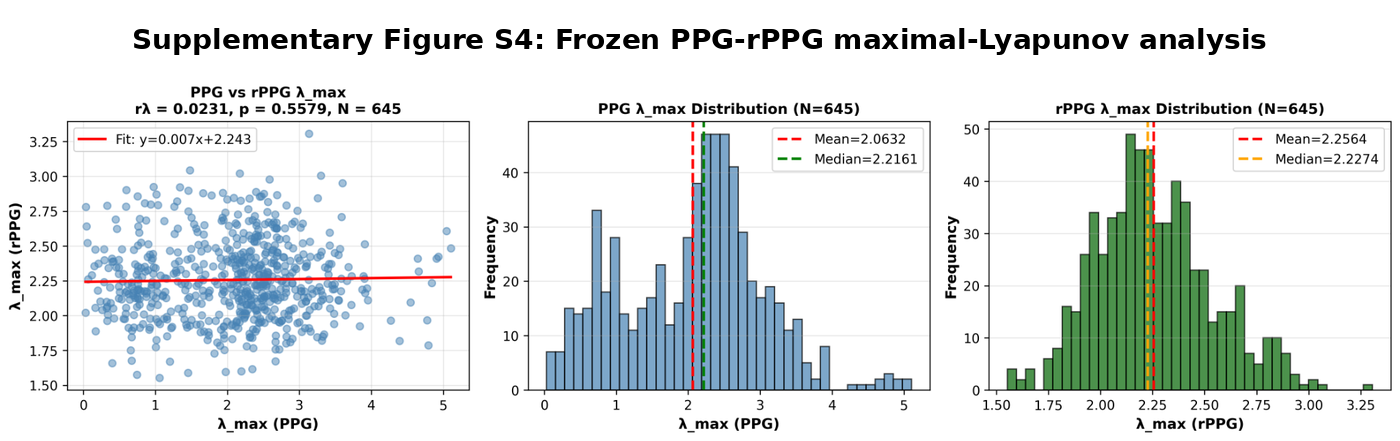}
\caption{\textbf{Frozen PPG--rPPG maximal-Lyapunov analysis.} \textit{The primary matched analysis used 645 recordings with valid PPG and rPPG $\lambda_{\max}$ estimates. No detectable matched recording-level correspondence was observed ($r_{\lambda}=0.0231$, 95\% CI -0.0542 to 0.1001; permutation $p=0.5584$), despite overlap in marginal values. The rPPG estimates also show marked variance compression relative to contact PPG, supporting the distinction between population-level similarity and preservation of recording-specific dynamical information.}}
\label{fig:s4}
\end{figure}

\begin{figure}[!htbp]
\centering
\includegraphics[width=0.82\textwidth]{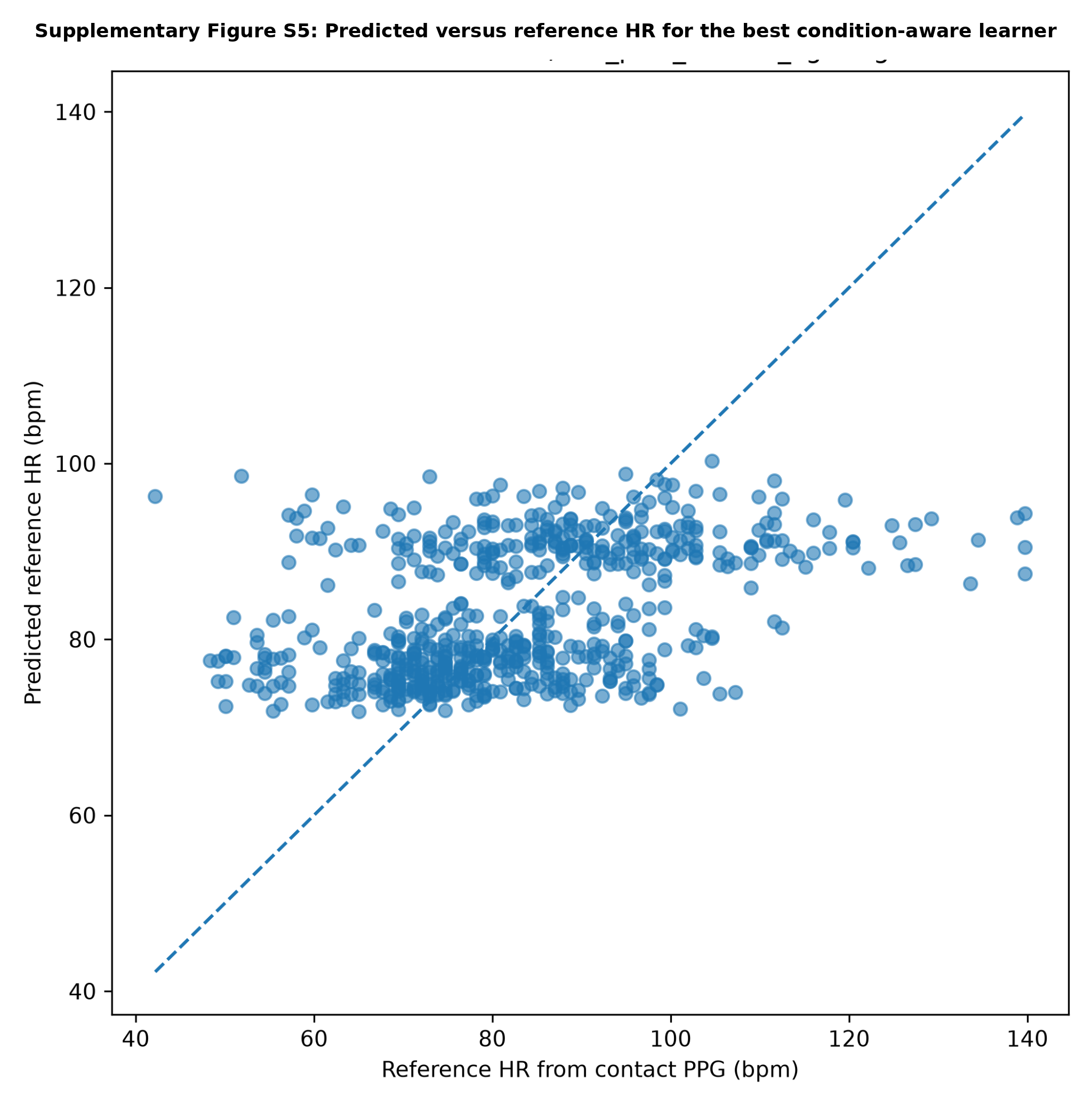}
\caption{\textbf{Predicted versus reference HR for the best condition-aware learner.} \textit{Out-of-fold predictions from the linear $M_2$ model (rPPG HR + motion + lighting) are plotted against synchronized contact-PPG reference HR. The identity line represents perfect recovery of contact-PPG HR. Although $M_2$ reduced MAE relative to the rPPG-only learner, predictions remain compressed toward central HR ranges and visibly depart from the identity line at the extremes, illustrating why the result is interpreted as partial gap reduction rather than PPG--rPPG equivalence. Predictions are strictly out-of-fold from subject-held-out validation.}}
\label{fig:s5}
\end{figure}

\begin{figure}[!htbp]
\centering
\includegraphics[width=\textwidth]{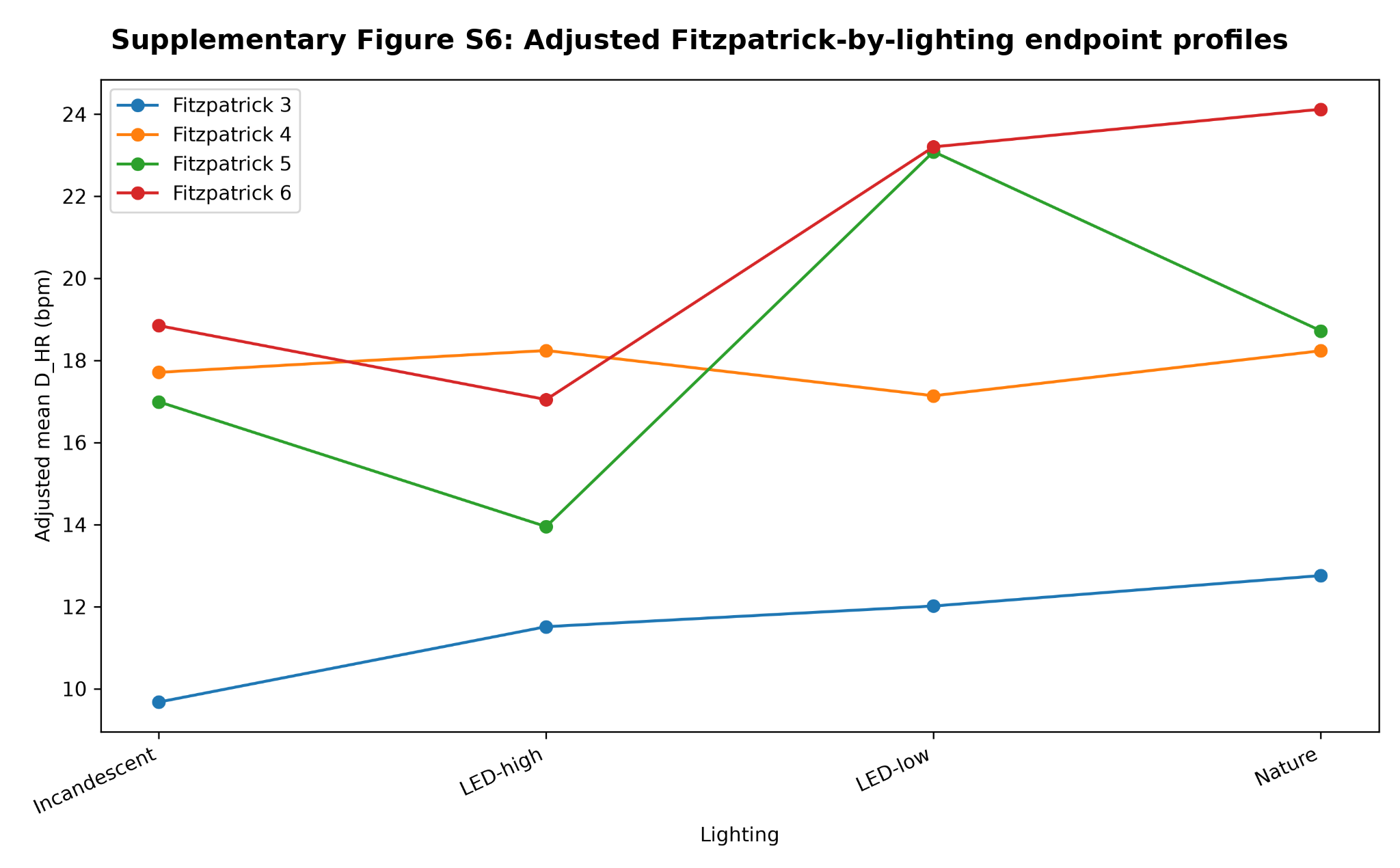}
\caption{\textbf{Adjusted Fitzpatrick-by-lighting endpoint profiles.} \textit{Model-based adjusted mean $D_{HR}$ is shown across lighting conditions for Fitzpatrick III--VI. Despite visible differences among cell means, the joint Fitzpatrick$\times$lighting block was not significant (Wald $\chi^2(9)=9.970$, $p=0.3529$); therefore, individual plotted differences are not interpreted as evidence of a compounded lighting-by-Fitzpatrick effect. The figure is retained to distinguish visually apparent subgroup variation from interaction heterogeneity supported by the repeated-measures model.}}
\label{fig:s6}
\end{figure}

The supplementary analyses provide additional validation and sensitivity checks for the main empirical findings. Supplementary Figure~\ref{fig:s1} provides the coefficient-focused Fitzpatrick endpoint gradient; Supplementary Figure~\ref{fig:s2} shows RGB-channel SNR distributions by Fitzpatrick group; Supplementary Figure~\ref{fig:s3} shows the SNR-adjusted coefficient sensitivity analysis; Supplementary Figure~\ref{fig:s4} shows the PPG--rPPG maximal-Lyapunov correspondence analysis; Supplementary Figure~\ref{fig:s5} shows out-of-fold predictions from the best condition-aware learner; and Supplementary Figure~\ref{fig:s6} shows the nonsignificant Fitzpatrick-by-lighting adjusted profiles.

\FloatBarrier

\section{Declarations}\label{declarations}

\subsection{Ethics statement}\label{ethics-statement}

This study is a secondary analysis of the existing de-identified Multi-Domain Mobile Video Physiology Dataset (MMPD) \cite{tang2023mmpd}. No new participants were recruited and no new data were collected for the present study.


\begin{thebibliography}{99}

\bibitem{allen2007ppg}
J. Allen, ``Photoplethysmography and its application in clinical physiological measurement,'' \emph{Physiological Measurement}, vol. 28, no. 3, pp. R1--R39, 2007. DOI: \url{https://doi.org/10.1088/0967-3334/28/3/R01}.

\bibitem{mcduff2015survey}
D. J. McDuff, J. R. Estepp, A. M. Piasecki, and E. B. Blackford, ``A survey of remote optical photoplethysmographic imaging methods,'' in \emph{2015 37th Annual International Conference of the IEEE Engineering in Medicine and Biology Society (EMBC)}, pp. 6398--6404, 2015. DOI: \url{https://doi.org/10.1109/EMBC.2015.7319857}.

\bibitem{wang2017algorithmic}
W. Wang, A. C. den Brinker, S. Stuijk, and G. de Haan, ``Algorithmic principles of remote PPG,'' \emph{IEEE Transactions on Biomedical Engineering}, vol. 64, no. 7, pp. 1479--1491, 2017. DOI: \url{https://doi.org/10.1109/TBME.2016.2609282}.

\bibitem{jacques2013optical}
S. L. Jacques, ``Optical properties of biological tissues: A review,'' \emph{Physics in Medicine and Biology}, vol. 58, no. 11, pp. R37--R61, 2013. DOI: \url{https://doi.org/10.1088/0031-9155/58/11/R37}.

\bibitem{dasari2021bias}
A. Dasari, S. K. A. Prakash, L. A. Jeni, and C. S. Tucker, ``Evaluation of biases in remote photoplethysmography methods,'' \emph{npj Digital Medicine}, vol. 4, p. 91, 2021. DOI: \url{https://doi.org/10.1038/s41746-021-00462-z}.

\bibitem{tang2023mmpd}
J. Tang, K. Chen, Y. Wang, Y. Shi, S. Patel, D. McDuff, and X. Liu, ``MMPD: Multi-Domain Mobile Video Physiology Dataset,'' in \emph{2023 45th Annual International Conference of the IEEE Engineering in Medicine \& Biology Society (EMBC)}, pp. 1--5, 2023. DOI: \url{https://doi.org/10.1109/EMBC40787.2023.10340857}.

\bibitem{takens1981detecting}
F. Takens, ``Detecting strange attractors in turbulence,'' in \emph{Dynamical Systems and Turbulence, Warwick 1980}, Springer, pp. 366--381, 1981. DOI: \url{https://doi.org/10.1007/BFb0091924}.

\bibitem{rosenstein1993practical}
M. T. Rosenstein, J. J. Collins, and C. J. De Luca, ``A practical method for calculating largest Lyapunov exponents from small data sets,'' \emph{Physica D: Nonlinear Phenomena}, vol. 65, nos. 1--2, pp. 117--134, 1993. DOI: \url{https://doi.org/10.1016/0167-2789(93)90009-P}.

\bibitem{he2022dynamics}
L. He, K. S. Alam, J. Ma, R. Povinelli, and S. I. Ahamed, ``Dynamics reconstruction of remote photoplethysmography,'' in \emph{Pervasive Computing Technologies for Healthcare: 15th EAI International Conference, PervasiveHealth 2021, Virtual Event, December 6--8, 2021, Proceedings}, Lecture Notes of the Institute for Computer Sciences, Social Informatics and Telecommunications Engineering, vol. 431, Springer, pp. 96--110, 2022. DOI: \url{https://doi.org/10.1007/978-3-030-99194-4_8}.

\bibitem{dehaan2013robust}
G. de Haan and V. Jeanne, ``Robust pulse rate from chrominance-based rPPG,'' \emph{IEEE Transactions on Biomedical Engineering}, vol. 60, no. 10, pp. 2878--2886, 2013. DOI: \url{https://doi.org/10.1109/TBME.2013.2266196}.

\bibitem{liu2023rppgtoolbox}
X. Liu, G. Narayanswamy, A. Paruchuri, X. Zhang, J. Tang, Y. Zhang, R. Sengupta, S. Patel, Y. Wang, and D. McDuff, ``rPPG-Toolbox: Deep remote PPG toolbox,'' \emph{Advances in Neural Information Processing Systems}, vol. 36, Datasets and Benchmarks Track, pp. 68485--68510, 2023. DOI: \url{https://doi.org/10.52202/075280-2995}.

\bibitem{setchfield2024skinoptics}
K. Setchfield, A. Gorman, A. H. R. W. Simpson, M. G. Somekh, and A. J. Wright, ``Effect of skin color on optical properties and the implications for medical optical technologies: A review,'' \emph{Journal of Biomedical Optics}, vol. 29, no. 1, p. 010901, 2024. DOI: \url{https://doi.org/10.1117/1.JBO.29.1.010901}.

\bibitem{weir2024skintone}
V. R. Weir, K. Dempsey, J. W. Gichoya, V. Rotemberg, and A.-K. I. Wong, ``A survey of skin tone assessment in prospective research,'' \emph{npj Digital Medicine}, vol. 7, p. 191, 2024. DOI: \url{https://doi.org/10.1038/s41746-024-01176-8}.

\bibitem{xiao2024review}
H. Xiao, T. Liu, Y. Sun, Y. Li, S. Zhao, and A. Avolio, ``Remote photoplethysmography for heart rate measurement: A review,'' \emph{Biomedical Signal Processing and Control}, vol. 88, Art. no. 105608, 2024. DOI: \url{https://doi.org/10.1016/j.bspc.2023.105608}.

\bibitem{elgendi2024sqi}
M. Elgendi, I. Martinelli, and C. Menon, ``Optimal signal quality index for remote photoplethysmogram sensing,'' \emph{npj Biosensing}, vol. 1, Art. no. 5, 2024. DOI: \url{https://doi.org/10.1038/s44328-024-00002-1}.



\bibitem{oladunni2025cfd}
T. Oladunni and A. Wong, ``Rethinking Multimodality: Optimizing Multimodal Deep Learning for Biomedical Signal Classification,'' \emph{IEEE Access}, vol. 13, pp. 156436--156464, 2025. DOI: \url{https://doi.org/10.1109/ACCESS.2025.3605315}.

\bibitem{oladunni2026cst}
T. Oladunni and F. G. Adewumi, ``Cardiac Stability Theory: An Axiomatically Grounded Framework for Continuous Cardiac Health Monitoring via Smartphone Photoplethysmography,'' arXiv:2604.23876, 2026. DOI: \url{https://doi.org/10.48550/arXiv.2604.23876}.


\bibitem{oladunni2026skin}
T. Oladunni and F. G. Adewumi,
``Skin-Tone-Robust Topological Signal Processing: A Framework for Bias-Reducing Optical Measurement Systems,''
\textit{medRxiv}, 2026, doi: 10.64898/2026.08.01.26359472.

\end{thebibliography}
\end{document}